\documentclass[preprint]{elsarticle}

\usepackage{amssymb}
\usepackage{graphicx}
\usepackage{caption}
\usepackage{pdfpages}
\usepackage{adjustbox}
\usepackage{rotating}
\usepackage{floatpag}
\usepackage{wrapfig}
\usepackage[T1]{fontenc}
\usepackage{mathrsfs }
\usepackage{multirow}
\usepackage{hhline}
\usepackage{enumerate}
\usepackage{enumitem}
\usepackage{adjustbox}
\usepackage[lined,commentsnumbered]{algorithm2e}
\usepackage{booktabs}
\usepackage{mathtools}

\usepackage{marvosym}

\biboptions{sort&compress}

\usepackage[english]{babel}
\usepackage[font=small,labelfont=bf]{caption}
\usepackage[tight,footnotesize]{subfigure}

\newlength\marincrease


\usepackage{makecell}

\usepackage{etoolbox}

\usepackage[bookmarks=true,
bookmarksnumbered=false, 
bookmarksopen=false, 
colorlinks=true,
linkcolor=webred]{hyperref}
\definecolor{webgreen}{rgb}{0, 0.5, 0} 
\definecolor{webblue}{rgb}{0, 0, 0.5} 
\definecolor{webred}{rgb}{0.5, 0, 0} 
\definecolor{webblack}{rgb}{0, 0, 0} 

\makeatletter

\let\old@@children\@@children
\def\@@children{\futurelet\my@next\my@@children}
\def\my@@children{%
\ifx\my@next\missing\else
\expandafter\@gobble
\fi
\expandafter\old@@children}

\makeatother

\newcommand{\missing}{ \edge[draw=none]; {} }

\journal{Elsevier}

\begin{document}

\renewcommand{\thesection}{\Roman{section}}

\renewcommand{\thesection}{\Roman{section}}
\newtheorem{theorem}{Theorem}
\newtheorem{proposition}{Proposition}
\newtheorem{eqed}{Example}
\newtheorem {lemmaa}{Lemma}
\newtheorem {observation}{Observation}
\newtheorem {example}{Example}
\newtheorem {corollary}{Corollary}
\newtheorem {defnn}{Definition}
\newenvironment{proof}{\noindent {\bf Proof :\ } }{\hfill$\Box$ }
\newtheorem {con}{Condition}
\newtheorem {conjecturee}{Conjecture}
\newtheorem {procd}{Procedure}
\newtheorem {rules}{Rule}

\newenvironment{proof of correctness}{\noindent {\bf Reason :\ } }{\hfill$\Box$ }
\newenvironment{definition}[1][Definition]{\begin{defnn} \sl}{\end{defnn}}
\newenvironment{algo}{\begin{algorithm}}{\end{algorithm}}

\begin{frontmatter}

\title{Analysis of Coronavirus Envelope Protein with Cellular Automata (CA) Model}

\author[mymainaddress1]{Raju Hazari\corref{mycorrespondingauthor}}
\cortext[mycorrespondingauthor]{Corresponding author}
\ead{hazariraju0201@gmail.com }

\address[mymainaddress1]{Department of Computer Science and Engineering, National Institute of Technology Calicut, Kerala, India 673601}

\author[mymainaddress2]{P Pal Chaudhuri}
\ead{ppc@carlbio.com}

\address[mymainaddress2]{Retired Professor, Indian Institute of Technology Kharagpur, India 721302}

\begin{abstract}
The reason of significantly higher transmissibility of SARS Covid (2019 CoV-2) compared to SARS Covid (2003 CoV) and MERS Covid (2012 MERS) can be attributed to mutations reported in structural proteins, and the role played by non-structural proteins (nsps) and accessory proteins (ORFs) for viral replication, assembly, and shedding. Envelope protein E is one of the four structural proteins of minimum length. Recent studies have confirmed critical role played by the envelope protein in the viral life cycle including assembly of virion exported from infected cell for its transmission. However, the determinants of the highly complex viral - host interactions of envelope protein, particularly with host Golgi complex, have not been adequately characterized. CoV-2 and CoV Envelope proteins of length 75 and 76 amino acids differ in four amino acid locations. The additional amino acid Gly (G) at location 70 makes CoV length 76. The amino acid pair EG at location 69-70 of CoV in place of amino acid R in location 69 of CoV-2, has been identified as a major determining factor in the current investigation. This paper concentrates on the design of computational model to compare the structure/function of wild and mutants of CoV-2 with wild and mutants of CoV in the functionally important region of the protein chain pair. We hypothesize that differences of CAML model parameter of CoV-2 and CoV characterize the deviation in structure and function of envelope proteins in respect of interaction of virus with host Golgi complex; and this difference gets reflected in the difference of their transmissibility. The hypothesis has been validated from single point mutational study on- (i) human HBB beta-globin hemoglobin protein associated with sickle cell anemia, (ii) mutants of  envelope protein of Covid-2 infected patients reported in recent publications. For each of these case studies, the CAML model parameter differ between wild and specific mutant that represent deviation of the structure-function of the mutant from that of wild leading to a specific disease. From detailed analysis of the characteristics of proteins of three viruses (CoV-2, CoV, MERS) and a few other virus proteins with CAML model, a Machine Learning (ML) framework has been designed. This ML framework enables us to report - (i) the contribution of higher transmissibility of CoV-2 envelope protein compared to CoV envelope protein, and (ii) a list of possible Covid-2 envelope protein mutants which may appear in future.
\end{abstract}


\end{frontmatter}

\section{Introduction}
\label{intro}

Seven different virus types, all with crown like structure associated with coronavirus disease (covid), were identified since 1965 causing common cold in human. Outbreak of the disease in 2003 in China, followed by the one in Middle East in 2012, brought the focus on the virus causing Severe Acute Respiratory Syndrome and so named as SARS Covid, and MERS Covid. Number of (infected persons, countries/region, death count) for 2003 SARS covid (CoV) and 2012 MERS covid are respectively (8000, 29, 774) and (2500, 27, 876). However, the novel virus SARS covid (CoV-2) identified in 2019 has resulted in the pandemic situation in 213 countries. It infected nearly 220 million persons with death count of more than 4.6 million. High speed train travel and air travel have increased significantly in last one decade leading to direct/indirect/close interaction among population across the globe; but that does not explain phenomenal increase of infected persons from a few thousands in earlier CoV to 220 million for CoV-2 till September 2021.                                                                                                                                                                                                              

The publications \cite{bianchi2020sars, sarkar2020structural, schoeman2020there, de2020improved, kuo2007exceptional, schoeman2019coronavirus, kuzmin2021structure, pervushin2009structure, hassan2020sars, wu2021effects, koyama2020variant, westerbeck2019infectious, lorizate2011role, maitra2020mutations, callaway2020making, jenna1998effect, cohen2011identification, v2021coronavirus, petersen2020comparing, cevik2020sars, kutter2018transmission, lee2010pdz, zhang2020viral, kawasuji2020transmissibility, venkatagopalan2015coronavirus, cai2018universal, hou2020sars, seyran2020structural, hassan2020possible, khailany2020genomic, loo1994functional, cabrera2021envelope}
highlight two vital issues - (a) entry of virus in host cell through binding of its spike protein with the receptor ACE2, followed by its replication in host cell, and (b) virus assembly/packaging through interaction with host (Golgi complex) prior to shedding. Four structural proteins of the virus are - Spike (S), Membrane (M), Nucleocapsid (N), and Envelope (E). The smallest size envelope protein plays a crucial role in the assembly of virus prior to secretion from infected cells \cite{bianchi2020sars, sarkar2020structural, schoeman2020there, de2020improved, kuo2007exceptional, schoeman2019coronavirus, kuzmin2021structure, pervushin2009structure, hassan2020sars, westerbeck2019infectious, lorizate2011role, cohen2011identification, v2021coronavirus, cevik2020sars, kutter2018transmission, zhang2020viral, kawasuji2020transmissibility, venkatagopalan2015coronavirus, kumar2021deletion, zheng2021tlr2, mukherjee2020host}. Some of the publications \cite{zheng2021tlr2} also pointed to the contribution of E protein for disease severity due to hyperactive cytokine release. Further, some authors \cite{v2021coronavirus, venkatagopalan2015coronavirus, cabrera2021envelope} emphasized the point that - the interaction of envelope protein with host Golgi complex has not been addressed adequately. Consequently, mutational study of envelope protein is a necessity to elucidate the deviation of a mutant from the wild in respect of its structure-function. CoV-2 envelope protein chain has 75 amino acids, while there are 76 amino acids for CoV with difference in four amino acid locations (55, 56, 69, and 70) covered by the C-terminal region \cite{hassan2020sars}. The amino acid pair EG at location 69-70 for CoV is replaced by single amino acid R for CoV-2. Experimental results of mutational study are reported for envelope protein of both Cov-2 (2019) and CoV (2003) with focus on the C-terminal region that covers Golgi complex targeting information \cite{venkatagopalan2015coronavirus} and conserved Motif of the chain pair. Comparison of the results shows high difference in CAML model parameters of CoV-2 wild and mutants compared to wild and mutants of CoV, with mutations inserted in same amino acid type in the functionally important region of the virus pair. In this background we propose the following hypothesis:

~difference of CAML model parameters of CoV-2 (wild, mutants) with

~CoV (wild, mutants) has a direct correlation with difference in the structure-

~function of envelope protein of two viruses. Assembly/packaging of virus

~prior to its shedding depends on the structure-function of C-terminal domain 

~of envelope protein interacting with host Golgi complex; consequently, 

~difference in structure-function points to the difference of transmissibility

~of CoV-2 and CoV/MERS. 

\noindent 
The hypothesis has been validated from mutational study on two proteins - (i) HBB beta-globin hemoglobin protein \cite{cai2018universal, NatureEducation2008, ncbiDatabase}, (ii) mutations reported in Covid-2 infected patients \cite{hassan2020sars, ncbiDatabase}. For each of these case studies, difference of CAML model parameter between wild and mutant corroborates the results reported in vitro/in vivo studies in respect of deviation of structure-function of specific mutant from its wild leading to a specific disease.

CAML model has been designed under the assumption that a mutation in amino acid chain of a protein originates as a point mutation at Coding DNA Sequence (CDS) that get transcripted to protein chain. Consequently, CAML model concentrates on the mutational study of both amino acid chain and CDS strand of Envelope Protein curated from NCBI \cite{ncbiDatabase}. CA preliminaries are introduced in Section~\ref{CA_preli}, followed by the CA model for amino acid of protein chain and nucleotide bases of CDS strand in Section~\ref{Sec:CA_rule_for_AA_CDS}.  Signal graphs derived out of evolution of the CA designed for wild and mutant amino acid chains and CDS strands are reported in Section~\ref{CA_Evolution_CL_Graph}; this section also reports CA model parameter derived out of signal graph analytics. A Machine Learning (ML) framework is next reported in Section~\ref{Sec:ML_Framework} to predict possible mutations in a protein for which structure-function of mutant differ from that of wild. The mutants identified by the ML framework cover the mutations reported with in vivo/in vitro studies for two case studies reported in Section~\ref{Three_Case_studies}. This section also presents a list of possible mutations in CoV-2 envelope protein which may appear in future. Experimental results for mutational study on Cov-2 and CoV (and also CoV-2 and MARS) are compiled and compared in Section~\ref{Comparison}.

\section{Cellular Automata (CA) Preliminaries}
\label{CA_preli} 

CA is a dynamical system discrete in space and time. The space is represented by a regular lattice in one, two or higher dimensions. Each site on the lattice also referred to as a cell, can be in one of a finite number of states. The next state of a cell depends on the current state of its neighbours and the associated next state function. Neumann \cite{von1996theory} proposed the CA model for constructing self-replicating machine employing 2000 cells, each holding 29 states. Subsequently, Wolfram \cite{wolfram1983statistical} proposed a simpler version of CA with two states per cell and three-neighbourhood. Research of a large number of authors from diverse disciplines has enriched the field of CA \cite{codd1968cellular, conway1970game, wolfram2002new, chaudhuri2018new, ppc1} and some researchers used CA to understand the dynamics of COVID-19 \cite{jithesh2021model, cavalcante2021modelling, schimit2021model, ghosh2020data, pereira2021deep}. The book \cite{ppc1} covers a comprehensive survey of CA theory and wide varieties of applications. Further, the universal appeal of CA model based on local interaction can be ascertained from the materials presented in the book \cite{ridley2015evolution} that highlights the effect of local (temporal and physical) changes transforming the human society from pre-historic to modern age.

\begin{figure}[h]
\begin{center}
     \includegraphics[scale=0.55]{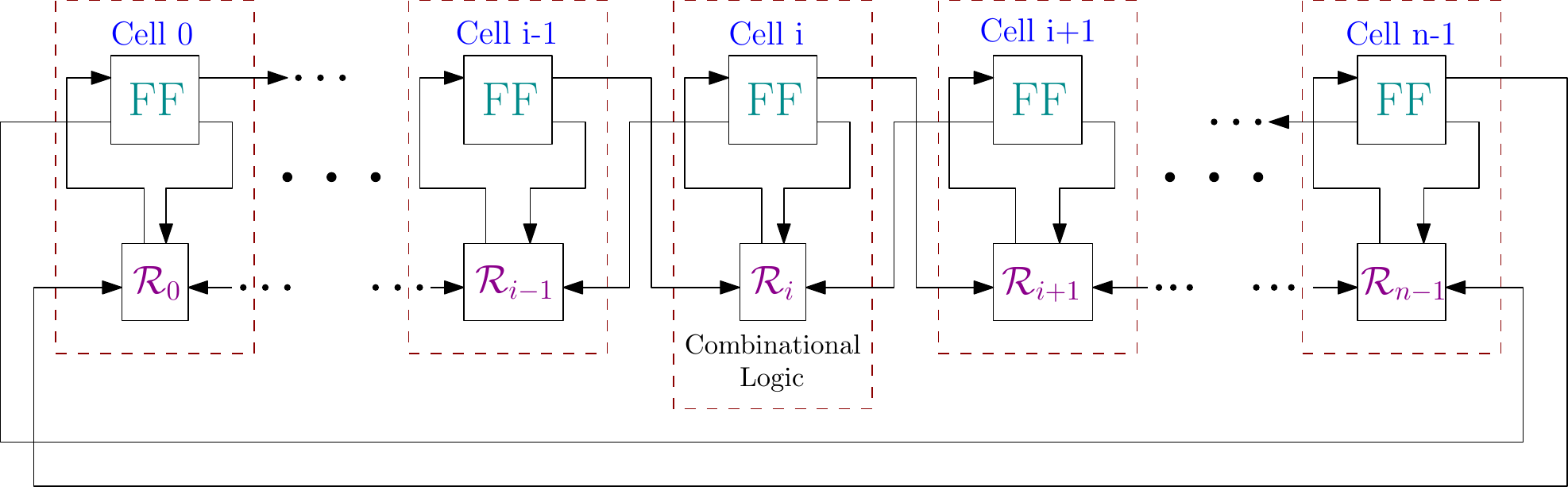}
     \caption{Implementation of an $n$-cell 3-neighborhood CA under periodic boundary condition}
     \label{Hardware-ECA-Periodic}
\end{center}
\end{figure}

\vspace{-3mm}
The CA model reported in this paper employs simplest CA structure with periodic boundary, three neighborhood, and two state per cell - referred to as 3NCA in the rest of the paper. The implementation of an $n$-cell 3-neighbourhood CA under periodic boundary condition is shown in Figure~\ref{Hardware-ECA-Periodic}. The middle cell is marked as $i^{th}$ cell while its left and right neighbors are denoted as $(i - 1)^{th}$ and $(i + 1)^{th}$ cell respectively. Consequently, $8(2^3)$ combinations exist for the triplet of current state values $<a_{i-1}  a_i  a_{i+1}>$ of $(i - 1)^{th}$ , $i^{th}$ and $(i + 1)^{th}$ cells. The next state function of a cell is shown in Figure~\ref{Hardware-ECA-Periodic}. The decimal value of the 8 bits of cell next state is referred to as `Rule' of evolution of a CA cell. There are $2^8 = 256$ such rules for three-neighborhood CA. The decimal value derived out of 8 bit string of two CA rules (45) and (105) are illustrated in Table~\ref{Trules} Row 4 and 5. Eight different input combinations of a CA rule is a triplet of binary bits - 111(7), 110(6), 101(5), 100(4), 011(3), 010(2), 001(1), 000(0), are noted on row 1 of Table~\ref{Trules}. Each combination represents the current state of $(i - 1)^{th}$, $i^{th}$, and $(i + 1)^{th}$ cells which can be viewed as three binary variables. In subsequent discussions we refer to the combinations of three input variables as Rule Min Terms (RMTs).  A RMT is referred to by its decimal value that varies from 0 to 7. Row 3 of Table~\ref{Trules} shows the cell next as $b_k$ ($b_k$ = 0 or 1), for $k$= 0 to 7. By convention, a CA rule is expressed as decimal counterpart of binary bit string $<b_7b_6b_5b_4b_3b_2 b_1b_0>$. So conventional weight assignment follows for conversion of 8-bit string of a rule to its decimal value with $w_7 = 2^7$, $w_6 = 2^6$ and so on. The decimal rule number 45 (00101101) is derived from its binary bit string as - $\sum_{k=0}^7 w_k b_k = 2^5 + 2^3 + 2^2 + 2^0 = 32 + 8 + 4 + 1 = 45$. A CA rule can be viewed as a Transform that accepts 3 bit input $k = <a_{i-1} a_i a_{i+1}>$ and generates single bit output $b_k$ in the next time of evolution of the cell. The cells of a 3NCA employs one of the 256 rules (0 to 255).

{\small
\begin{table*}[h]
\[
\begin{array}{|cccccccccc|}
\hline
{\rm Present~ State:} &  111 & 110 & 101 & 100 & 011 &  010 &  001 &  000 & Rule \\ 
(RMT)	& (7) & (6) & (5) & (4) & (3) & (2) & (1) & (0) & \\ 
{\rm  Next~ State ~(b_k)}    & b_7  & b_6 & b_5 & b_4 & b_3  & b_2  & b_1  & b_0  &  \\ \hline
  {\rm ~ (i)~ Next~ State:}    &   0  &  0 &  1  &  0  &   1  & 1  &   0  &   1  & 45 \\
{\rm (ii)~ Next~ State:}    & 0  &  1 &  1  &  0  &   1  & 0  &   0  &   1   & 105 \\
\hline
\end{array}
\]
\vspace{-3mm}
\caption{Look-up table for rule 45 and 105}
\label{Trules}
\end{table*}
}

In order to design rules to build CA model for a physical system, it is convenient to use the following format for a rule.  We employ this format for design of rules for 20 amino acids and 4 nucleotide bases in next section.

\vspace{2mm}
\noindent
\textbf{1-major and 0-major (7653 4210) Format:} Out of 8 RMTs, the binary string of each of the 4 RMTs (7, 6, 5, 3) has two 1's, while RMTs (4, 2, 1, 0) has two 0's - these two classes are  referred to as 1-Major  and 0-Major RMTs respectively in Table~\ref{ECA_rule_Table} that shows rules 45 and 105 in this format.

\begin{table}[h]
\centering
\resizebox{0.95\textwidth}{!}{
\begin{tabular}{|c|cccc|cccc|}
\hline 
\multirow{2}{*}{Rule} & \multicolumn{4}{c|}{1-Major RMTs} & \multicolumn{4}{c|}{0-Major RMTs} \\ 
 & 7(111) & 6(110) & 5(101) & 3(011) & 4(100) & 2(010) & 1(001) & 0(000) \\ 
\hline 
45 & 0 & 0 & 1 & 1 & 0 & 1 & 0 & 1 \\ 
\hline 
105 & 0 & 1 & 1 & 1 & 0 & 0 & 0 & 1 \\ 
\hline 
\end{tabular} }
\caption{Representation of two CA rules 45 and 105 in 1-major and 0-major format}
\label{ECA_rule_Table}
\end{table}

\begin{figure}[h]
\hfill
\subfigure[Ten cells  of a CA configured with ten rules \label{ECARules}]{\includegraphics[width=5cm, height=8mm]{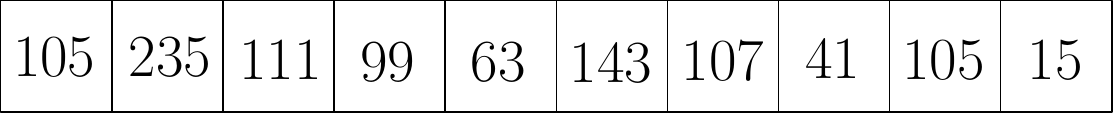}}
\hfill
\subfigure[Rule Vector\label{Rule-Vector}]{\includegraphics[width=5cm, height=8mm]{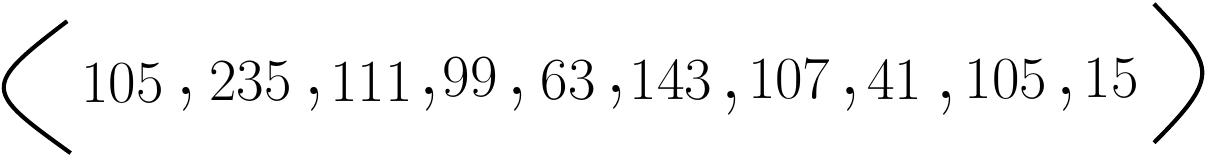}}
\hfill
\caption{Rule vector for a 10 cell 3NCA}
\end{figure}

\section{CA Rule Vector for Amino Acid and CDS strand of a Protein}
\label{Sec:CA_rule_for_AA_CDS}

Design of CA model for a physical system involves design of rule for each CA cell. A cell models the smallest component of the system so that the physical domain features of the component get represented in the binary bit pattern of its rule. Once the rule design step is complete, we can represent a 3NCA as a Rule Vector - a string of 3NCA rules (0 to 255). Figure~\ref{Rule-Vector} shows the rule vector for a 10 cell CA with the rules specified in Figure~\ref{ECARules}. The CA evolves at each time step through local interaction of each cell with its neighborhood. The design proceeds through a number of iterations to build a model that characterizes the system with efficient mapping of model parameters derived out of CA evolution to the physical domain features of the system. The string of 3NCA Rule Vector can be employed to model the biological strings - DNA, RNA, Codon string, and Amino Acid chain of a protein.

\begin{figure}[h]
\begin{center}
\includegraphics[width=12cm, height=8.5cm]{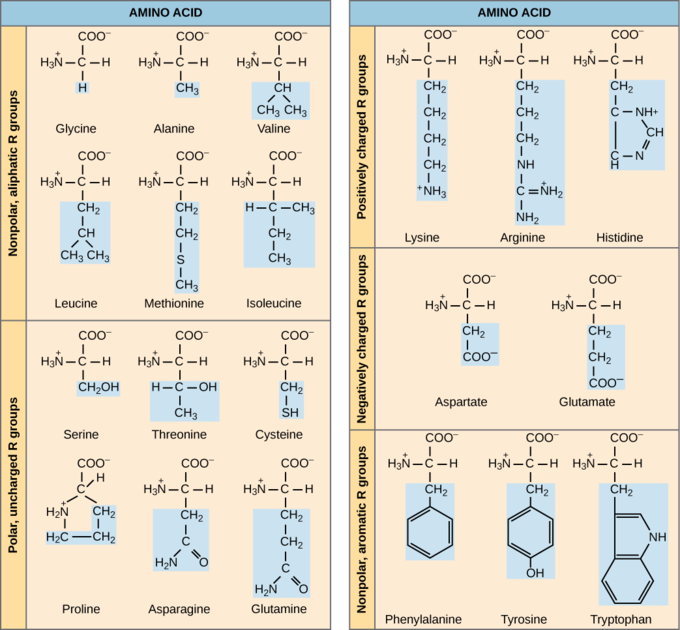}
\caption{Atomic structure of 20 common amino acids(AA) divided into 5 groups}
\label{20_AAcide}
\end{center}
\end{figure}

The CA model for 20 amino acids has been derived from the first principle by considering the atomic structure of amino acid molecules. Figure~\ref{20_AAcide} shows atomic structure of 20 amino acids divided into 5 different groups based on the property of the side chain referred to as Residue (R). The side chain property - non-polar, polar, positively charge, negatively charged, aromatic - is defined by the structure of interconnected atoms. For design of CA rules, the atoms are divided into two groups -  H-atom and non-H atom (Carbon C, Nitrogen N, Oxygen O, Sulphur S), each having different proton count in nucleus and electron count in outer shells. The CA model for hydrogen atom H, lightest one in the periodic table elements, treated differently from non-H atoms.

\subsection{Design of CA rule for Amino Acid (AA) Chain of Protein}

\vspace{2mm}
Each amino acid has a common Back-Bone with 9 atoms - 5 non-H atoms (2 C, 2 O, 1 N) and 4 H atoms. The side chain residue (R) have widely different number of atoms - 1 atom (for Gly - G) to 18 atoms (for Arg - R). CA rule for amino acid is designed as a composite module of backbone connected to side chain. This is a departure from the convention of referring to an amino acid as a residue. In the rest of the paper we refer to an amino acid (AA) as a composite molecule covering both backbone and residue.

\subsubsection{Design Methodology}
\label{Design_meth_AA}

\vspace{2mm}
3NCA rules for amino acid molecules are defined as a sequence of rules (decimal value 0 to 255). On considering the atomic structure of amino acids, it is convenient to design the rules by considering the RMT string <7 6 5 4 3 2 1 0> of a rule in the form of 1-Major and 0-Major RMTs as <7653   4210>(see Table~\ref{ECA_rule_Table}). The H atoms are assigned as `1' in the next state of 0-major RMTs, while non-H atoms are assigned as `1' in the next state of 1-major RMTs. However, we do not populate more than 6 number of `1's in the 8-bit pattern of a rule. Hence, rather than a single rule, the design employs a string of rules if the molecule has more than 4 H atoms or non-H atoms.

\subsubsection{Design of CA rule for Amino Acid Backbone}

\vspace{2mm}
As per this design methodology, a string of two CA rules designed for 9 atom amino acid  backbone as <107 99>is reported in Table~\ref{AA_backbone}, where the first column shows design of two rules expressed in the 1-Major 0-Major format for 9 atoms (4 H atoms and 5 non-H atoms). The fourth column presents two rules in the <7 6 5 4 3 2 1 0> format; finally, column 5 shows the rule string with rules expressed in decimal value. Thus, the CA backbone is modelled with two CA cells configured with a string of two CA rules <107 99>.

\begin{table}[h]
\begin{center}
\resizebox{\textwidth}{!}{ 
\renewcommand{\arraystretch}{1.5} 
\begin{tabular}{|c|c|c|c|c|}
\hline 
Rules for AA backbone in & Number  & Number of   & Rules for AA Backbone in & Rule string for \\  
<7653 4210> format & of & atoms in (AA)  & <7 6 5 4 3 2 1 0> format &  AA backbone \\                                                                                                                                                                                                                                                             
                   &   rules    & backbone         &                      & in Decimal  \\             
\hline 
<0111 0011><0110 0011> & 2 & 2C-N-2O-4H(9) & <0110 1011><0110 0011> & <107 99> \\ 
\hline 
\end{tabular} }
\caption{Design of CA rule string for amino acid backbone having 9 atoms - two carbon, one nitrogen, two oxygen, and 4 hydrogen atoms}
\label{AA_backbone}
\end{center}
\end{table}

\vspace{-8mm}
\subsubsection{Design of CA rule for Amino Acid (AA)}
\label{CA_rule_forAA}

\vspace{2mm}
The atomic structure of 20 amino acid are presented in Figure~\ref{20_AAcide}. Following the design methodology mention in Section~\ref{Design_meth_AA}, rule strings for 20 AA are designed, which are reported in Table~\ref{AA_Composite_rule_Vector}. The first column presents the design of rule string expressed in 1-major 0-major <7653 4210>format with column 2, 3, 4 respectively showing - number of rules, AA name, number of atoms (H and non-H) in the side chain. The rule string is shown in column 5 in the <7 6 5 4 3 2 1 0> format  and the corresponding decimal value (0 to 255)  in column 6. Finally, the composite rules for an amino acid with backbone and side chain rule concatenated is reported in column 7 of Table~\ref{AA_Composite_rule_Vector}. For example, the composite Rule string <107 99><63 11 15> in column 7 is designed for the amino acid Ile (I) with 9 atoms for common backbone and 13 atoms for side chain shown in $6^{th}$ row proceeds as follows. First a string of two rules <107 99> is designed for backbone (Table~\ref{AA_backbone}). A string of 3 rules is next designed for 13 atoms of side chain with 9 H atoms and 4 non-H atoms. Intermediate columns 5 and 6 show the rules expressed in RMT format <7 6 5 4 3 2 1 0> and their decimal counterpart. Last four rows of the Table~\ref{AA_Composite_rule_Vector} display the nucleotide base triplet for 20 amino acids and 3 stop codons. In the rest of the paper, an amino acid is referred to as AA modeled with 3 to 5 cells configured with a string of 3 to 5 rules. The AAs are serially marked as 0, 1, 2 -- with met (M) as the $0^{th}$ AA of an amino acid chain of a protein.

\begin{table}[h]
\centering
\resizebox{1.0\textwidth}{!}{ 
\renewcommand{\arraystretch}{1.5}
{\bf
\begin{tabular}{|c|c|c|c|c|c|c|}
\hline 
Rules for AA side chain in  & No. of  & Amino acid & No. of & Rules for AA side chain in & Rule string for AA  & Composite Rule String  \\ 
<7653 4210>format & rules & (AA) name & atoms & <7 6 5 4 3 2 1 0> format & side chain in Decimal & for Amino acid \\ 
\hline 
<0000 0001> & 1 & Gly(G) & H(1) & <0000 0001> & <1> & <107 99><1> \\ 
\hline 
<0001 0111> & 1 & Ala(A) & CH3(4) & <0000 1111> & <15> & <107 99><15> \\ 
\hline 
<0011 1111><0001 0111> & 2 & Val(V) & C3H7(10) & <0011 1111><0000 1111> & <59 15> & <107 99><59 15> \\ 
\hline 
<0001 0011><0001 0111> & 3 & Leu(L) & C4H9(13) & <0000 1011><0000 1111> & <11 15 63> & <107 99><11 15 63> \\ 
<0011 1111> &   &  &  & <0011 1111> &  &  \\
\hline 
<0011 1111><0011 0111> & 2 & Met(M) & SC3H7(11) & <0011 1111><0011 0111> & <63 47> & <107 99><63 47> \\ 
\hline 
<0011 1111><0001 0011> & 3 & Ile(I) & C4H9(13) & <0011 1111><0000 1011> & <63 11 15> & <107 99><63 11 15> \\
 <0001 0111> &   &  &  & <0000 1111> &  &  \\ 
\hline 
<0011 0111> & 1 & Ser(S) & COH3(5) & <0010 1111> & <47> & <107 99><47> \\ 
\hline 
<0011 0011><0001 0111> & 2 & Thr(T) & C2OH5(8) & <0010 1011><0000 1111> & <43 15> & <107 99><43 15> \\ 
\hline 
<1001 0111> & 1 & Cys(C) & SCH3(5) & <1000 1111> & <143> & <107 99><143> \\ 
\hline 
<0001 0011><0011 1111> & 2 & Pro(P) & C3H6(9) & <0000 1011><0011 1111> & <11 63> & <107 99><11 63> \\ 
\hline 
<0001 0011><0111 0011> & 2 & Asn(N) & C2NOH4(8) & <0000 1011><0110 1011> & <11 107> & <107 99><11 107> \\ 
\hline 
<0011 1111><0111 0011> & 2 & Gln(Q) & C3NOH6(11) & <0011 1111><0110 1011> & <63 107> & <107 99><63 107> \\ 
\hline 
<0011 1111><0011 1111> & 3 & Lys(K) & C4NH11(16) & <0011 1111><0011 1111> & <63 63 15> & <107 99><63 63 15> \\ 
 <0001 0111> &   &  &  & <0000 1111> &  &  \\
\hline 
<0011 1111><0111 0111> & 3 & Arg(R) & C4N3H11(18) & <0011 1111><0110 1111> & <63 111 63> & <107 99><63 111 63> \\ 
 <0011 1111> &   &  &  & <0011 1111> &  &  \\
\hline 
<0111 0111><0111 0011> & 2 & His(H) & C4N2H5(11) & <0110 1111><0110 1011> & <111 107> & <107 99><111 107> \\ 
\hline 
<1111 0011> & 1 & Asp(D) & C2O2H2(6) & <1110 1011> & <235> & <107 99><235> \\ 
\hline 
<0001 0011><1111 0011> & 2 & Glu(E) & C3O2H4(9) & <0000 1011><1110 1011> & <11 235> & <107 99><11 235> \\ 
\hline 
<0001 0011><0111 0011> & 3 & Phe(F) & C7H7(14) & <0000 1011><0110 1111> & <11 107 111> & <107 99><11 107 111> \\ 
 <0111 0111> &   &  &  & <0110 1111> &  &  \\
\hline 
<0001 0011><0111 0111> & 3 & Tyr(Y) & C7OH7(15) & <0000 1011><0110 1111> & <11 111 111> & <107 99><11 111 111> \\ 
 <0111 0111> &   &  &  & <0110 1111> &  &  \\
\hline 
<0111 0111><1111 0011> & 3 & Trp(W) & C9NH8(18) & <0110 1111><1110 1011> & <111 239 111> & <107 99><111 239 111> \\ 
 <0111 0111> &   &  &  & <0110 1111> &  &  \\
\hline 
\end{tabular} } }
\caption{CA rule string for AA residue (R) \& Concatenated Composite Rule String (backbone \& R) for AA}
\label{AA_Composite_rule_Vector}
\end{table}

\subsubsection{CA Rule Vector for AA Chain}

\vspace{2mm}
On completion of rule string design for each amino acid, rule vector can be derived for the AA chain of a protein by concatenating rule strings of each AA. Table~\ref{AA_Chain} shows Envelope Protein AA string for - (a) CoV-2 (2019) - 75 AA, (b) CoV (2003) - 76 AA, and (c) Hemoglobin subunit beta HBB Protein - 147 AA; partial rule vectors for three AA chains are illustrated.

\begin{table}
\centering
\resizebox{1.0\textwidth}{!}{ 
\renewcommand{\arraystretch}{1.2}
\begin{tabular}{|c|l|}
\hline 
\multicolumn{2}{|c|}{\textbf{CoV-2 (2019) - 75 AA}} \\ 
\hline 
AA Chain & mysfvseetgtlivnsvllflafvvfllvtlailtalrlcayccnivnvslvkpsfyvysrvknlnssrvpdllv  \\ 
\hline 
Partial & <107, 99, 11, 111, 111, 107, 99, 47, 107, 99, 11, 107, 111, 107, 99, 59, 15,  \\  
rule vector & 107, 99, 47, 107, 99, 11, 235, $\cdots$ > \\ 
\hline 
\multicolumn{2}{|c|}{\textbf{CoV (2003) - 76 AA}} \\  
\hline 
AA Chain & mysfvseetgtlivnsvllflafvvfllvtlailtalrlcayccnivnvslvkptvyvysrvknlnssegvpdllv \\ 
\hline 
Partial & <107, 99, 11, 111, 111, 107, 99, 47, 107, 99, 11, 107, 111, 107, 99, 59, 15, \\ 
rule vector & 107, 99, 47, 107, 99, 11, 235, $\cdots$ > \\ 
\hline 
\multicolumn{2}{|c|}{\textbf{HBB Hemoglobin Protein - 147 AA}} \\  
\hline 
\multirow{3}{*}{AA Chain} & mvhltpeeksavtalwgkvnvdevggealgrllvvypwtqrffesfgdlstpdavmgnpkvkahgkk \\  
 & vlgafsdglahldnlkgtfatlselhcdklhvdpenfrllgnvlvcvlahhfgkeftppvqaayqkvvagv \\
 & analahkyh \\ 
\hline 
Partial & <107, 99, 59, 15, 107, 99, 111, 107, 107, 99, 11, 15, 63, 107, 99, 43, 15, \\ 
rule vector & 107, 99, 11, 63, 107, 99, 11, 235, $\cdots$ > \\ 
\hline 
\end{tabular} }
\caption{Amino acid chain and partial (first 6 excluding Met(M)) rule vector for - (a) Wild CoV-2 (2019) -75 AA, (b) Wild CoV (2003) - 76 AA, and (c) Wild Hemoglobin subunit beta - 147 AA}
\label{AA_Chain}
\end{table}

Once the CA Rule Vector is designed for a protein, we study the evolution of the CA generating a Signal Graph referred to as Cycle Length Signal Graph (CL Signal Graph) detailed in Section~\ref{CA_Evolution_CL_Graph} following CA rule design for nucleotide bases of CDS strand of double helix DNA string.

\subsection{CA Rule Vector for CDS Strand }

\begin{figure}[h!]
\begin{center}
\includegraphics[width=11cm, height=3.7cm]{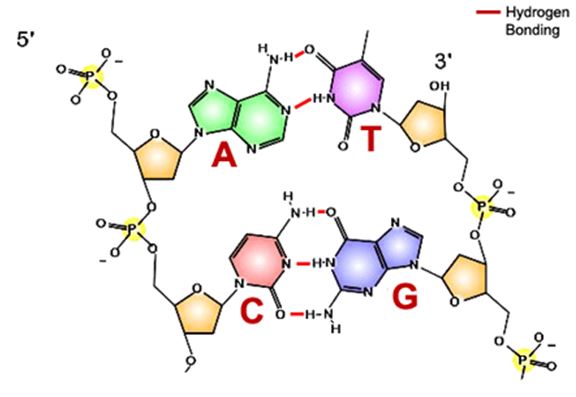}
\caption{Backbone connected to nucleotide bases of DNA string and its complimentary version forming double helix structure}
\label{CDS_Strand}
\end{center}
\end{figure}

\vspace{2mm}
Figure~\ref{CDS_Strand} shows the DNA sugar-phosphate backbone connected to nucleotide bases. As per the design methodology noted in Section~\ref{Design_meth_AA}, rule strings for backbone and nucleotide bases are designed and reported in Table~\ref{CDS_CA_Rule}. Column 2 shows the atom count for the molecule noted in column 1. Rule string in <7653  4210> and  <7 6 5 4 3 2 1 0> format are noted in column 3 and 4, while column 5 reports the rule string in decimal value. The composite rule string of six rules for common backbone and four bases are noted in column 6. Table~\ref{CDS_Chain_rule_vector} shows the CDS strand \cite{ncbiDatabase} and partial CA rule vector for - (a) CoV-2 Envelope Protein, (b) CoV Envelope Protein, and (c) HBB Hemoglobin protein along with partial rule vector for 1st six amino acids excluding $1^{st}$ AA Met(M) with start codon atg.

\begin{table}[h]
\centering
\resizebox{1.0\textwidth}{!}{ 
\renewcommand{\arraystretch}{1.5}
{\bf
\begin{tabular}{|c|c|c|c|c|c|}
\hline 
\multirow{2}{*}{Molecule} & Atom count & Rule string in & Rule string in & Rule string in & <backbone><base>  \\ 
 & (Non-H + H) & <7653 4210> format & <7654 3210> format & decimal & string \\ 
\hline 
DNA  & (PC5O5H7) & <1111 0011><1111 0011> & <1110 1011><1110 1011> & <235 235 111> & \multirow{2}{*}{-} \\ 
 Backbone  & (11+7) & <0111 0111> & <0110 1111> &   &  \\
\hline 
Cytosine(C) & (C4N3OH4) & <0011 0001><0011 0001> & <0010 1001><0010 1001> & <41 41 235> & <235 235 111><41 41 235> \\ 
  & (8+4) & <1111 0011> & <1110 1001> &   &  \\
\hline 
Guanine(G) & (C5N5OH4) & <0111 0001><1111 0001> & <0110 1001><1110 1001> & <105 233 235> & <235 235 111><105 233 235> \\ 
  & (11 + 4) & <1111 0011> & <1110 1011> &   &  \\
\hline 
Thiamine(T) & (C5N2O2H5) & <1111 0001><0111 0011> & <1110 1001><0110 1011> & <233 107 43> & <235 235 111><233 107 43> \\
  & (9 + 5) & <0011 0011> & <0010 1011> &   &  \\ 
\hline 
Adenine(A) & (C5N5H4) & <1111 0001><1111 0001> & <1110 1001><1110 1001> & <233 233 43> & <235 235 111><233 233 43> \\ 
  & (10 + 4) & <0011 0011> & <0010 1011> &   &  \\
\hline 
\end{tabular} } }
\caption{3NCA rule string for backbone and nucleotide bases C, G, T, A}
\label{CDS_CA_Rule}
\end{table}

\begin{table}[h!]
\centering
\resizebox{1.0\textwidth}{!}{ 
\renewcommand{\arraystretch}{1.4}
\begin{tabular}{|c|l|}
\hline 
\multicolumn{2}{|c|}{\textbf{CoV-2 (2019) - 75 AA}} \\  
\hline 
 &  atgtactcattcgtttcggaagagacaggtacgttaatagttaatagcgtacttctttttcttgctttcgtggtattc  \vspace{-2mm} \\ 
CDS Strand & ttgctagttacactagccatccttactgcgcttcgattgtgtgcgtactgctgcaatattgttaacgtgagtcttgta \vspace{-2mm} \\ 
 &  aaaccttctttttacgtttactctcgtgttaaaaatctgaattcttctagagttcctgatcttctggtc   \\   
\hline 
 & \small{<235, 235, 111, 233, 107, 43, 235, 235, 111, 233, 233, 43, 235, 235, 111, 41, 41, 235, 235,} \vspace{-2mm} \\
 & \small{235, 111, 233, 107, 43, 235, 235, 111, 41, 41, 235, 235, 235, 111, 233, 233, 43, 235, 235,} \vspace{-2mm}   \\  
Partial & \small{111, 233, 107, 43, 235, 235, 111, 233, 107, 43, 235, 235, 111, 41, 41, 235, 235, 235, 111,} \vspace{-2mm} \\  
rule vector & \small{105, 233, 235, 235, 235, 111, 233, 107, 43, 235, 235, 111, 233, 107, 43, 235, 235, 111, 233,} \vspace{-2mm}  \\  
 & \small{107, 43, 235, 235, 111, 41, 41, 235, 235, 235, 111, 105, 233, 235, 235, 235, 111, 105, 233,} \vspace{-2mm} \\
 & \small{235, 235, 235, 111, 233, 233, 43, 235, 235, 111, 233, 233, 43, $\cdots$ >}    \\  
\hline
\multicolumn{2}{|c|}{\textbf{CoV (2003) - 76 AA}} \\  
\hline 
 & atgtactcattcgtttcagaagaaacaggtacgttaatagttaatagcgtacttctcttcttggctttcgtggtatt \vspace{-2mm} \\  
CDS Strand & cttgctagtcacactagccatccttactgcgcttcgattgtgtgcgtactgctgcaatattgttaacgtgagtttgg \vspace{-2mm} \\ 
 & taaaaccaacagtttacgtttactcacgtgttaaaaatctgaactcttctgagggagttcctgatcttctggtc  \\
\hline 
 & \small{<235, 235, 111, 233, 107, 43, 235, 235, 111, 233, 233, 43, 235, 235, 111, 41, 41, 235, 235,} \vspace{-2mm} \\
 & \small{235, 111, 233, 107, 43, 235, 235, 111, 41, 41, 235, 235, 235, 111, 233, 233, 43, 235, 235,} \vspace{-2mm} \\  
 &  \small{111, 233, 107, 43, 235, 235, 111, 233, 107, 43, 235, 235, 111, 41, 41, 235, 235, 235, 111,} \vspace{-2mm} \\  
Partial & \small{105, 233, 235, 235, 235, 111, 233, 107, 43, 235, 235, 111, 233, 107, 43, 235, 235, 111,} \vspace{-2mm} \\ 
rule vector & \small{233, 107, 43, 235, 235, 111, 41, 41, 235,  235, 235, 111, 233, 233, 43, 235, 235, 111, 105,}  \vspace{-2mm}\\  
 & \small{233, 235, 235, 235, 111, 233, 233, 43, 235, 235, 111, 233, 233, 43, $\cdots$ >} \\
\hline 
\multicolumn{2}{|c|}{\textbf{HBB Hemoglobin Protein - 147 AA}} \\  
\hline 
\multirow{4}{*}{CDS Strand} & 				  atggtgcatctgactcctgaggagaagtctgccgttactgccctgtggggcaaggtgaacgtggatgaagttggt \vspace{-2mm} \\
 & ggtgaggccctgggcaggctgctggtggtctacccttggacccagaggttctttgagtcctttggggatctgtcca \vspace{-2mm} \\
 & ctcctgatgctgttatgggcaaccctaaggtgaaggctcatggcaagaaagtgctcggtgcctttagtgatggcct \vspace{-2mm} \\  
 & ggctcacctggacaacctcaagggcacctttgccacactgagtgagctgcactgtgacaagctgcacgtggatcct \vspace{-2mm} \\ 
 & gagaacttcaggctcctgggcaacgtgctggtctgtgtgctggcccatcactttggcaaagaattcaccccaccagt \vspace{-2mm} \\ 
 & gcaggctgcctatcagaaagtggtggctggtgtggctaatgccctggcccacaagtatcac \\ 
\hline 
 & \small{<235, 235, 111, 105, 233, 235, 235, 235, 111, 233, 107, 43, 235, 235, 111, 105, 233, 235,} \vspace{-2mm}  \\ 
 & \small{235, 235, 111, 41, 41, 235, 235, 235, 111, 233, 233, 43, 235, 235, 111, 233, 107, 43, 235,} \vspace{-2mm}  \\ 
 & \small{235, 111, 41, 41, 235, 235, 235, 111, 233, 107, 43, 235, 235, 111, 105, 233, 235, 235,} \vspace{-2mm} \\  
Partial & \small{235, 111, 233, 233, 43, 235, 235, 111, 41, 41, 235, 235, 235, 111, 233, 107, 43, 235, 235,} \vspace{-2mm}  \\  
rule vector & \small{111, 41, 41, 235, 235, 235, 111, 41, 41, 235, 235, 235, 111, 233, 107, 43, 235, 235, 111,} \vspace{-2mm} \\ 
 & \small{105, 233, 235, 235, 235, 111, 233, 233, 43, 235, 235, 111, 105, 233, 235, $\cdots$ >} \\
\hline  
\end{tabular} }
\caption{Wild CDS strand and partial (18 bases for first 6 amino acids excluding Met(M)) CA rule vector of three AA chain of - (a) CoV-2 Envelope Protein, (b) CoV Envelope Protein, and (c) HBB Hemoglobin Protein noted in Table~\ref{AA_Chain}}
\label{CDS_Chain_rule_vector}
\end{table}

\section{CA Evolution Generating Cycle Length Signal Graph (CL Signal Graph) for AA Chain and CDS Strand of Protein}
\label{CA_Evolution_CL_Graph}

The CA model employs periodic boundary 3NCA (Figure~\ref{Hardware-ECA-Periodic}) for an amino acid chain of a protein excluding first amino acid Met (M) of the chain. The cells of an n cell CA are serially marked as $0, 1, 2, \cdots (n-1)$, where $0^{th}$ cell and $(n-1)^{th}$ are neighbors of each other. An amino acid (backbone and side chain) is modelled with a string of 3 to 5 cells configured with 3 to 5 rules, as reported in Table~\ref{AA_Composite_rule_Vector}. The CA evolves in successive time steps as per the Rule Vector that specifies the rules assigned for each CA cell. The cells of the CA are initialized with an alternating sequence of 1 and 0. The evolution continues till each cell settles down in a cycle, where the cell cycles through a set of binary states; the length of the cycle is referred to as \emph{Cycle Length (CL)}. The CL signal graph refers to the graph showing CL values for each cell on y-axis, with x-axis displaying serial location of CA cells. If a few cells do not settle down in a cycle even after 4000-time steps, these are marked with value CL(-5).

\subsection{CL Signal Graph Analytics for Wild Type AA Chain and Mutant}
\label{CL_Signal_Graph_Analytics_AA}

\begin{figure}[h]
\begin{center}
\includegraphics[width=13.3cm, height=4.0cm]{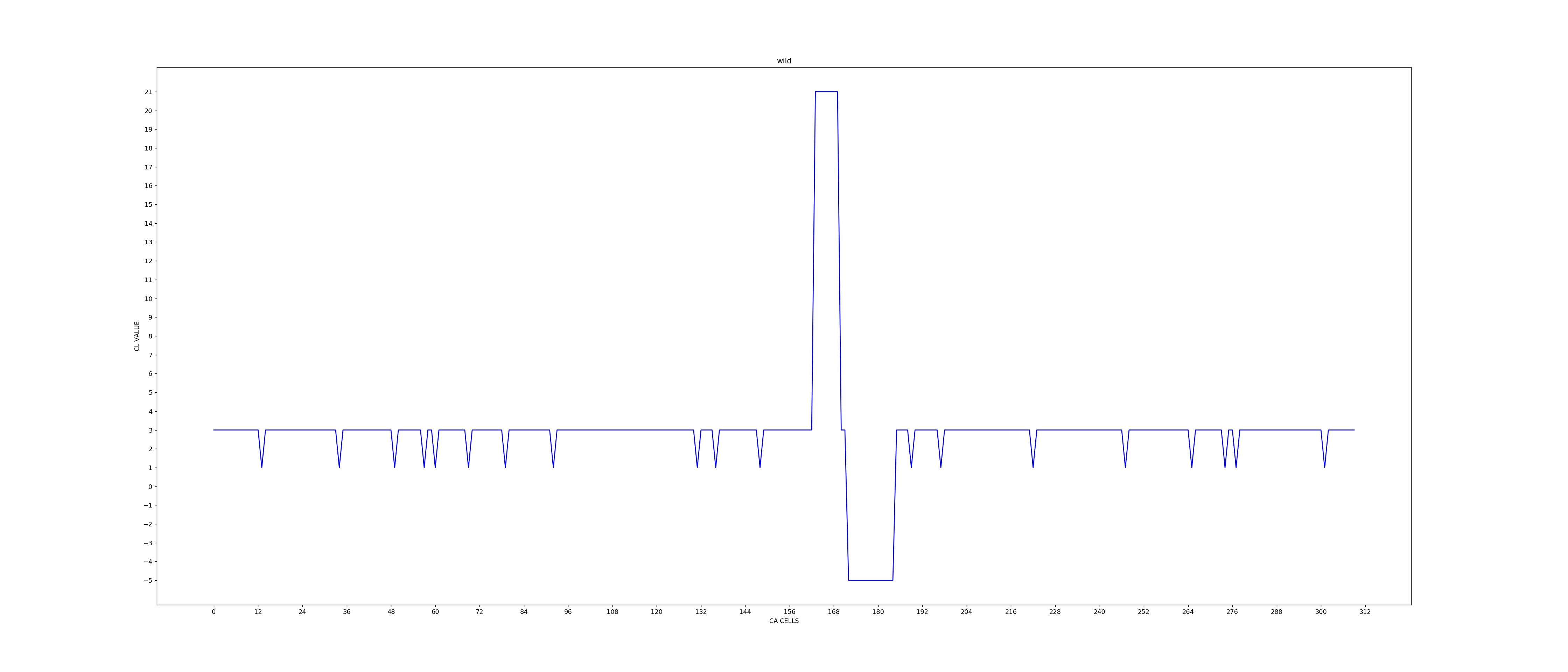}
\caption{CL signal graph for wild envelope protein of CoV-2 AA chain}
\label{CL_Graph_Wild_AA_Cov2}
\end{center}
\end{figure}

\begin{figure}[h]
\begin{center}
\includegraphics[width=13.3cm, height=4.9cm]{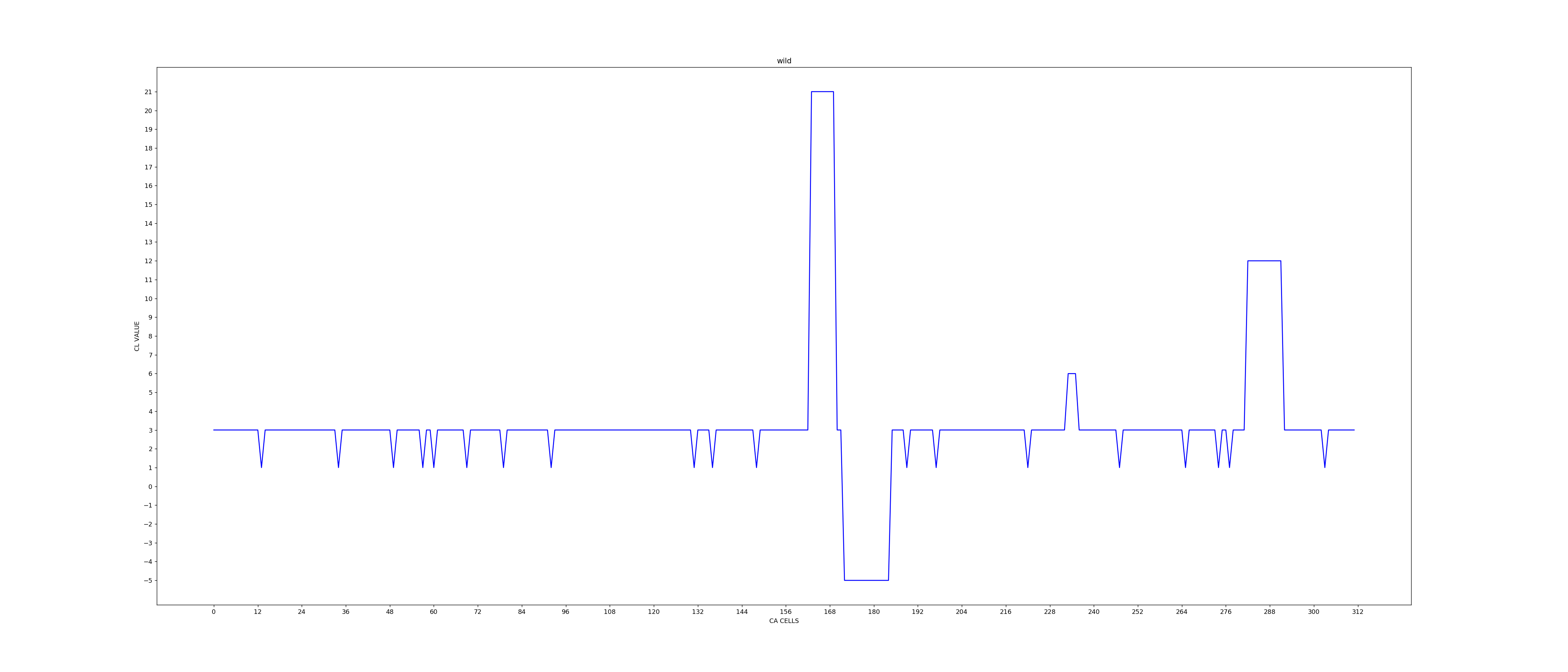}
\caption{CL signal graph for wild envelope protein of CoV AA chain}
\label{CL_Graph_Wild_AA_Cov}
\end{center}
\end{figure}

\begin{figure}[h!]
\begin{center}
\includegraphics[width=13.3cm, height=4.9cm]{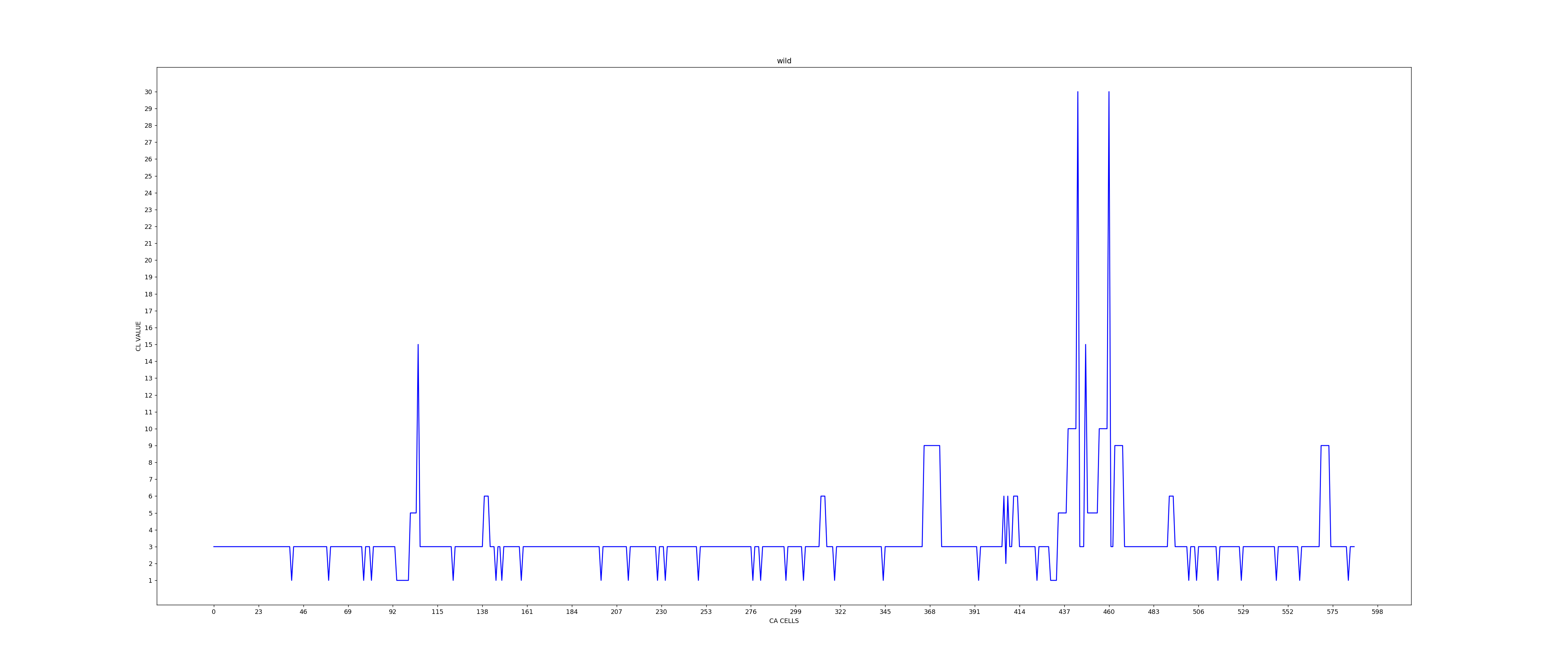}
\caption{CL signal graph for wild HBB hemoglobin protein AA chain}
\label{CL_Graph_Wild_AA_HBB}
\end{center}
\end{figure}

\vspace{2mm}
For the CAML model, signal graph analytics enable extraction of meaningful information from a graph relevant for mutational study of protein. \emph{Wild type AA chain} refers to the one that can be observed in nature without any mutation. The chain with a mutation is termed as \emph{Mutant}. CL signal graphs for wild type AA chain are shown in Figure~\ref{CL_Graph_Wild_AA_Cov2}, \ref{CL_Graph_Wild_AA_Cov} and \ref{CL_Graph_Wild_AA_HBB}. In view of common backbone, the base line of the CL signal graph has the CL value 3. Signal graph analytics proceeds on identifying \emph{mcl} (Maximum Cycle Length) and \emph{N-mcl} (Next to mcl) signals. The mcl signal is the one that covers a number of cells with maximum cycle length value. More than one mcl signals may exist in a CL graph. A signal with CL value less than mcl but greater than y is referred to as N-mcl, where number of signals with CL values (y+1) and y is greater than 3. Such an evaluation of N-mcl signals enables avoidance of noisy signals below the CL value (y+1) for signal graph analytics. The value of the parameter y will depend on the analysis of low valued signals in a CL signal graph.

\subsubsection{CL Signal Graph for Wild Type AA Chain}

\vspace{2mm}
Figure~\ref{CL_Graph_Wild_AA_Cov2}, \ref{CL_Graph_Wild_AA_Cov} and \ref{CL_Graph_Wild_AA_HBB} shows the CL signal graphs for evolution of CA as per rule vector for wild type AA chain noted in Table~\ref{AA_Chain} for CoV-2, CoV Envelope Protein, and HBB Hemoglobin Protein respectively.

\subsubsection{CL Signal Graph for Mutant AA Chain}

\begin{figure}[h!]
\begin{center}
\includegraphics[width=13.3cm, height=4.7cm]{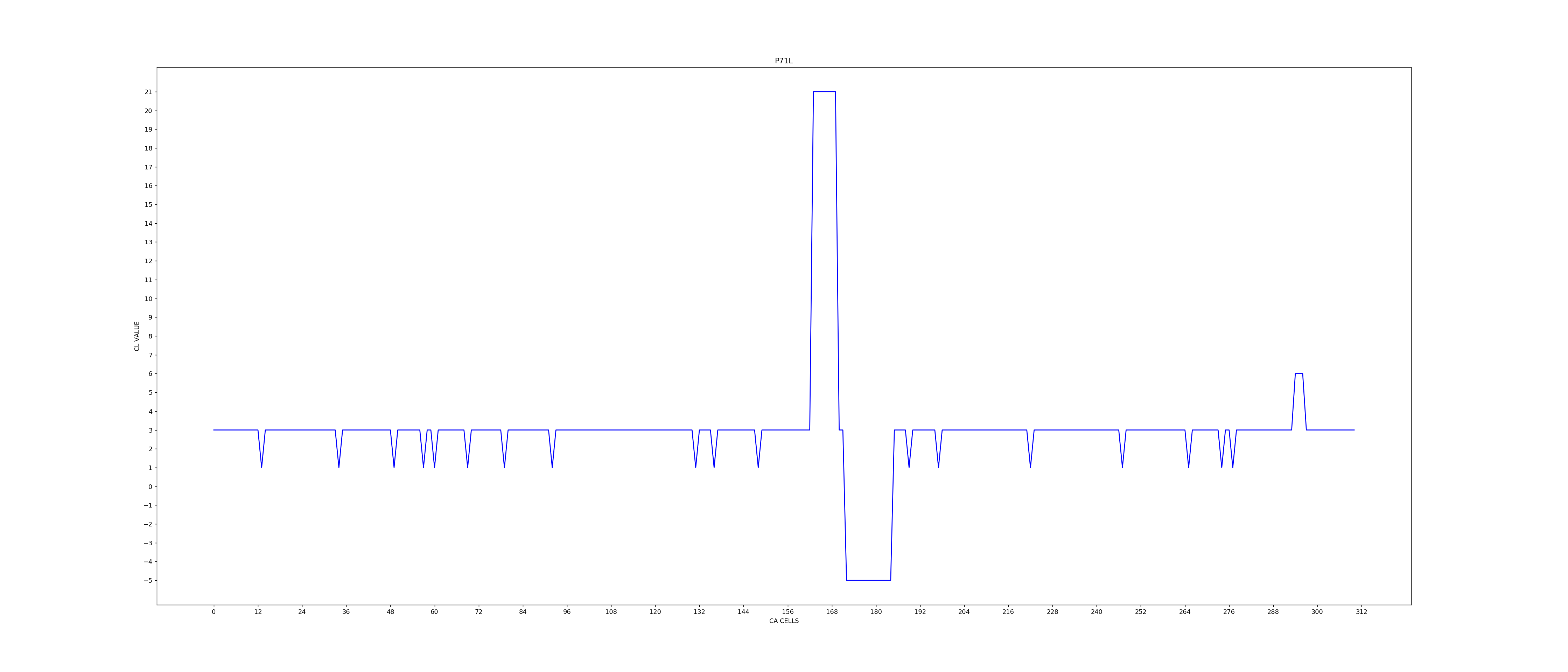}
\caption{CL Signal graph for mutant P71L for CoV-2 envelope protein}
\label{CL_Graph_Cov2_AA_Mutant_P71L}
\end{center}
\end{figure}

\begin{figure}[h!]
\begin{center}
\includegraphics[width=13.3cm, height=4.7cm]{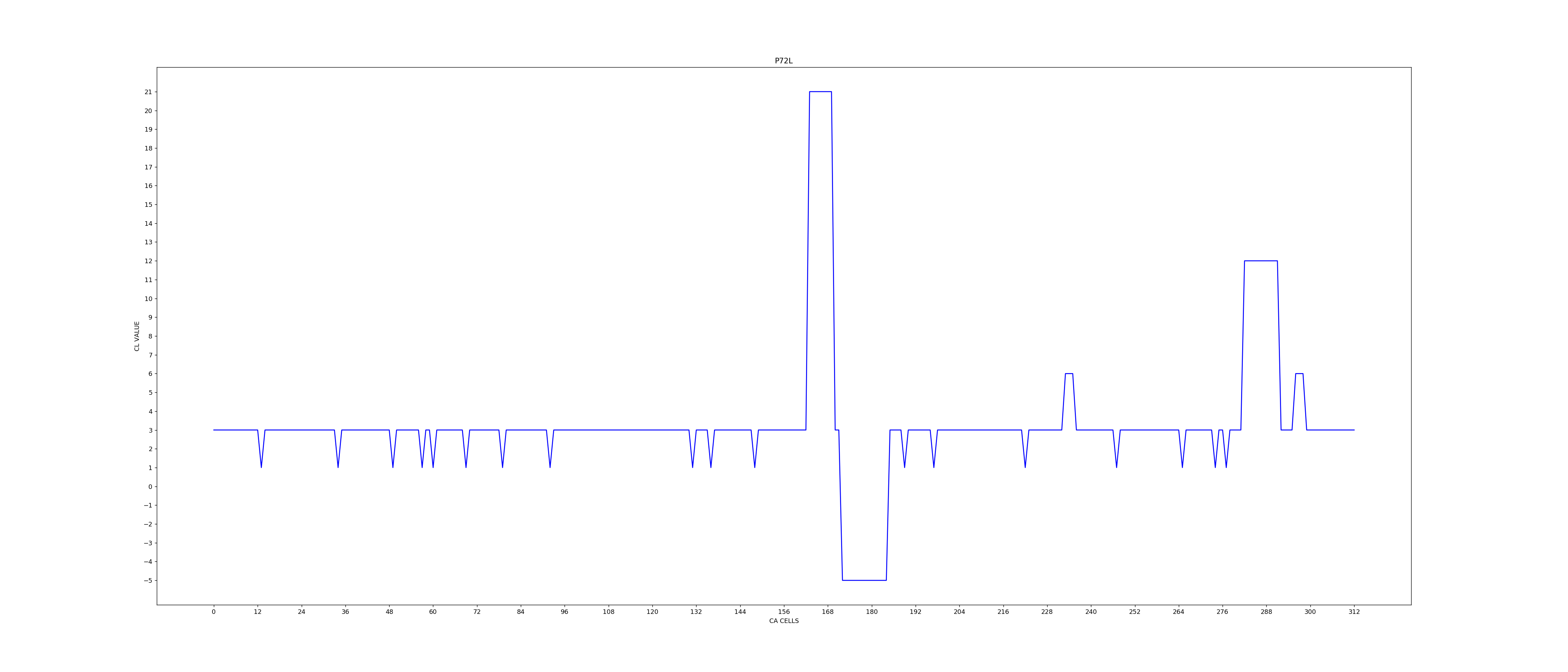}
\caption{CL Signal graph for mutant P72L for CoV envelope protein}
\label{CL_Graph_Cov_AA_Mutant_P72L}
\end{center}
\end{figure}

\begin{figure}[h!]
\begin{center}
\includegraphics[width=13.3cm, height=4.7cm]{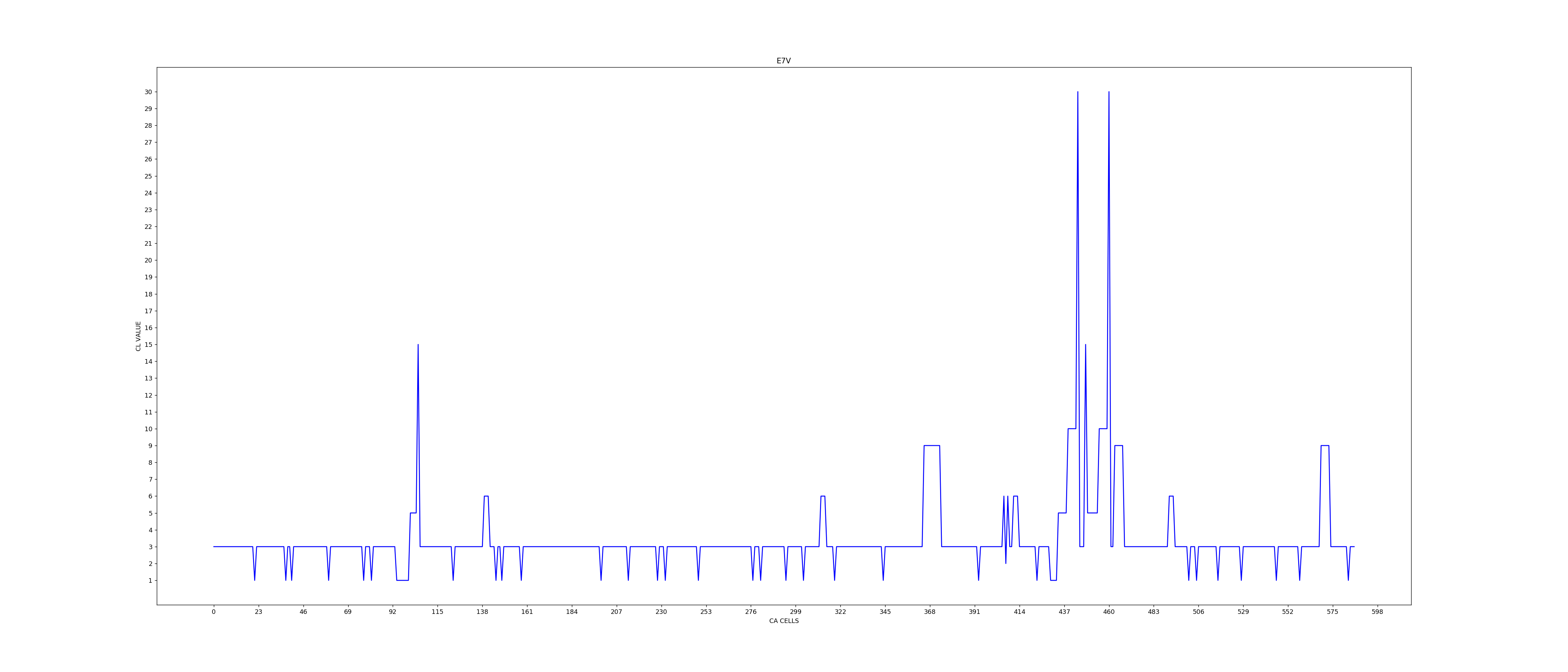}
\caption{CL Signal graph for mutant E7V for HBB Hemoglobin protein}
\label{CL_Graph_HBB_AA_Mutant_E7V}
\end{center}
\end{figure}

\vspace{2mm}
Figure~\ref{CL_Graph_Cov2_AA_Mutant_P71L}, \ref{CL_Graph_Cov_AA_Mutant_P72L} and \ref{CL_Graph_HBB_AA_Mutant_E7V} display CL signal graph for three mutants - (i) P71L (original amino acid Proline (P) at serial location 71 mutated with amino acid Leu (L)) for CoV-2 Envelope Protein AA chain; (ii) P72L (original amino acid Proline (P) at serial location 72 mutated with amino acid Leu (L)) for CoV Envelope Protein AA chain; (iii) E7V (original amino acid Glu (E ) at serial location 7 mutated with amino acid Val (V)) for HBB Hemoglobin protein.

\subsubsection{Evaluation of Difference of CL Value Sum (DCLVS-AA) between Wild and Mutant AA Chain Signal Graphs}
\label{Evaluation_of_Difference}

\vspace{2mm}
CL Value Sum (CLVS) for a wild is computed considering mcl and Nmcl signals of wild CL graph. Similarly, CLVS is computed out of a mutant CL graph. Next, we derive: 

\vspace{2mm}
{\small{\emph{DCLVS-AA (Difference of CL Value Sum-AA) = Difference in CLVS of Wild and mutant CL signal graph for AA chain}}}

\vspace{2mm}
\noindent
This CA model parameter DCLVS-AA, as reported in Section~\ref{Comparison}, is also employed to compare difference in structure-function of a pair of envelope proteins of (CoV-2, CoV, MERS) with similar length and function; length is assumed to be similar if difference in length is less than 10\% of total chain length. The following three categories of cells are excluded for evaluation of CLVS-AA: \vspace{2mm}

\noindent
\emph{(i)Cells not displaying cycle} : For a CL signal graph of AA chain, there may exist (say $N^{'}$ cells) which do not reach a cycle; these $N^{'}$ cells are marked with CL(-5) value.

\vspace{2mm}
\noindent
\emph{(ii)Cells displaying noise signal with CL value less than (y+1), as explained in Section~\ref{CL_Signal_Graph_Analytics_AA}} : Signals with CL value less than (y+1) are assumed to be noise in the CL signal graph for AA chain. For the current version (y+1) = 6. \vspace{2mm}

\noindent
\emph{(iii)Cells in a split Nmcl signal} : A split Nmcl signal displays multiple peaks with the cells between peaks have CL value less than (y+1) = 6. Figure~\ref{CL_Graph_Wild_AA_MERS} illustrates a CL signal graph displaying two Nmcl signals at cell locations (93 to 100) and (121 to 129), each with two peaks. Since CL value between two peaks is 5, we ignore such Nmcl signals for computation of CLVS.

\subsection{CL Signal Graph for Wild Type and Mutant CDS strand}

\begin{figure}[h!]
\begin{center}
\includegraphics[width=13.3cm, height=4.1cm]{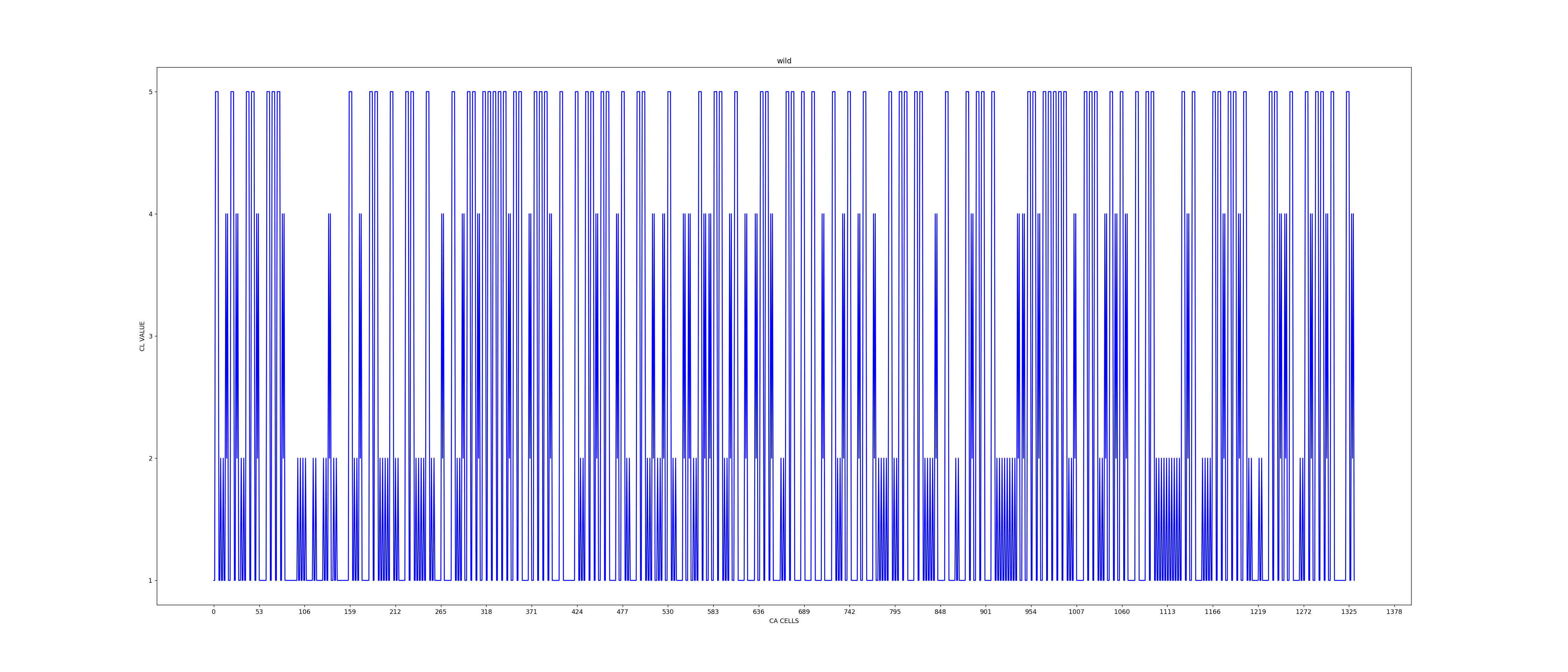}
\caption{CL signal graph for CDS wild strand for CoV-2 envelope protein}
\label{CL_Graph_Wild_CDS_Cov2}
\end{center}
\end{figure}

\begin{figure}[h!]
\begin{center}
\includegraphics[width=13.3cm, height=4.1cm]{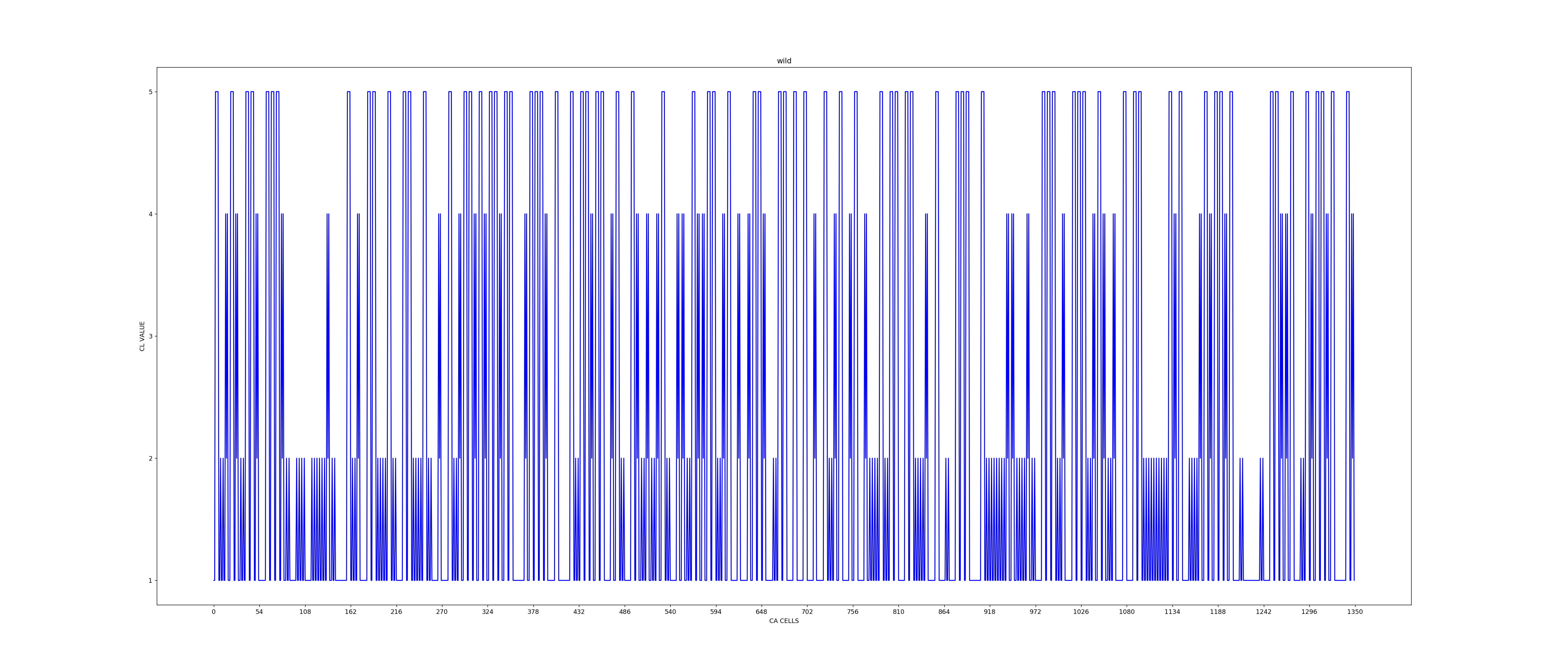}
\caption{CL signal graph for CDS wild strand for CoV envelope protein}
\label{CL_Graph_Wild_CDS_Cov}
\end{center}
\end{figure}

\begin{figure}[h!]
\begin{center}
\includegraphics[width=13.3cm, height=4.1cm]{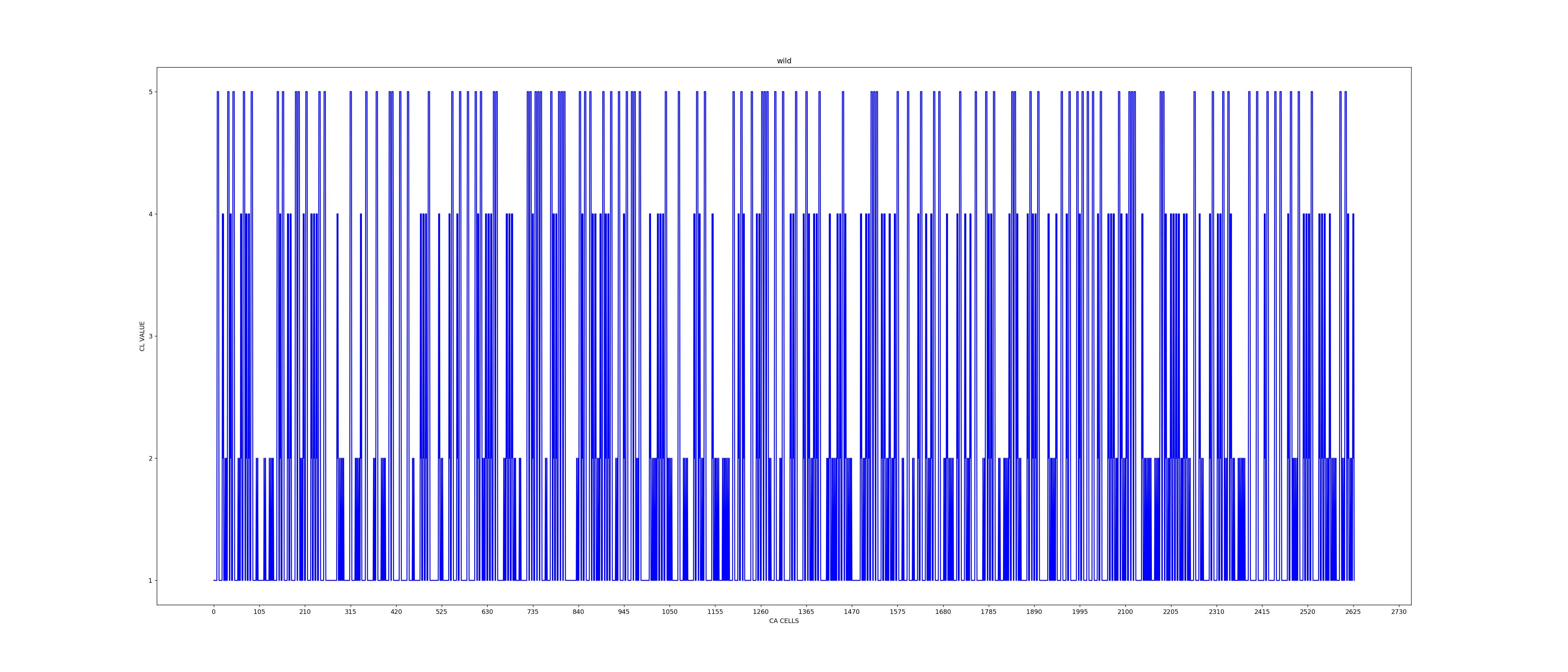}
\caption{CL signal graph for CDS wild strand for HBB Hemoglobin protein}
\label{CL_Graph_Wild_CDS_HBB}
\end{center}
\end{figure}

\vspace{2mm}
Each of the four nucleotide bases of CDS strand is modelled with six rules - a string of three rules for sugar phosphate backbone and three rules for bases attached to the backbone (Table~\ref{CDS_CA_Rule}). In view of common backbone, the CL signal graph for a CDS strand shows five signal values 1, 2, 4, 5, 10. Table~\ref{CDS_Chain_rule_vector} reports the CDS strand and partial rule vector for AA chain of three proteins. Figure~\ref{CL_Graph_Wild_CDS_Cov2}, \ref{CL_Graph_Wild_CDS_Cov} and \ref{CL_Graph_Wild_CDS_HBB} report the CL signal graph for evolution of CA as per the rule vector for CDS Wild strand noted in Table~\ref{CDS_Chain_rule_vector} for CoV-2 Envelope Protein, CoV Envelope Protein, and HBB hemoglobin protein.

\begin{figure}[h]
\begin{center}
\includegraphics[width=13.3cm, height=4.3cm]{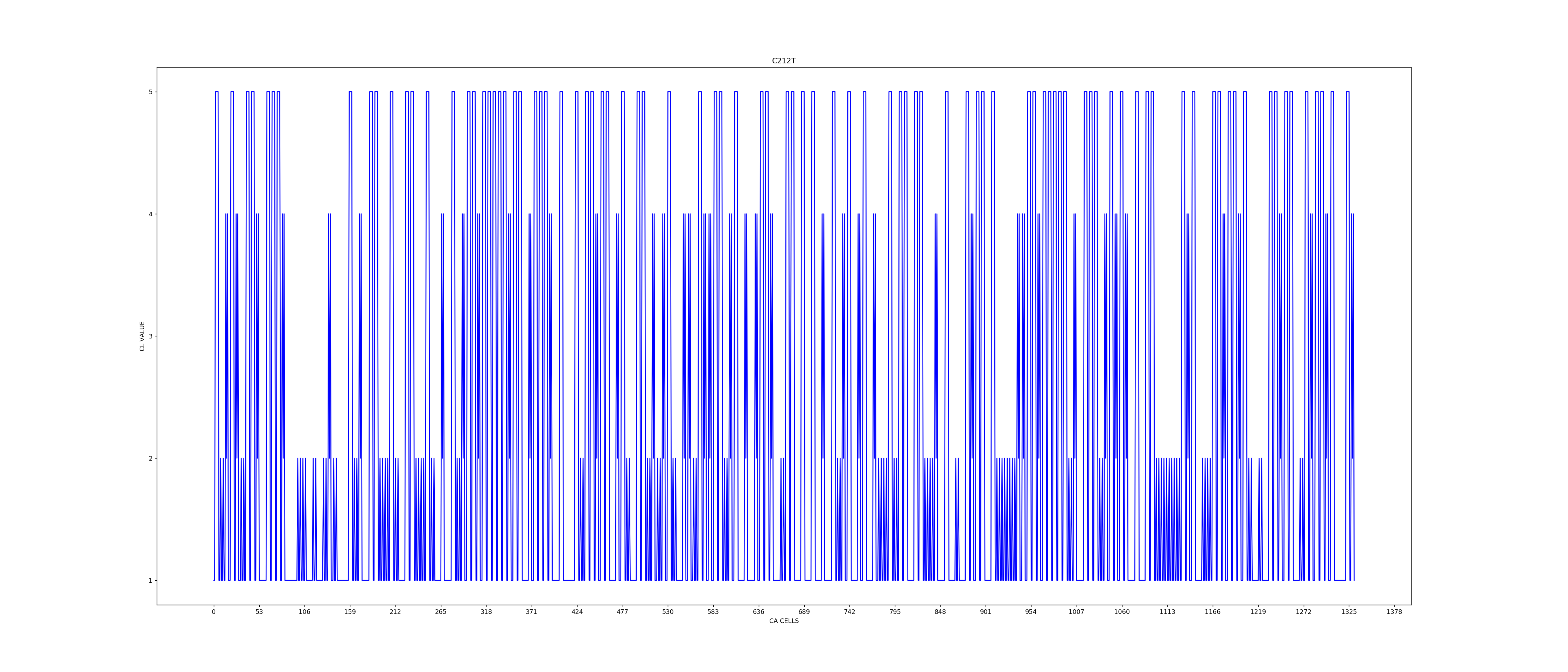}
\caption{Partial CL signal graph for mutant CDS strand with mutation c26402t for CoV-2 envelope protein chain (corresponds to P71L at AA level)}
\label{CL_Graph_Cov2_CDS_Mutant_c2640t}
\end{center}
\end{figure}

\begin{figure}[h!]
\begin{center}
\includegraphics[width=13.3cm, height=4.3cm]{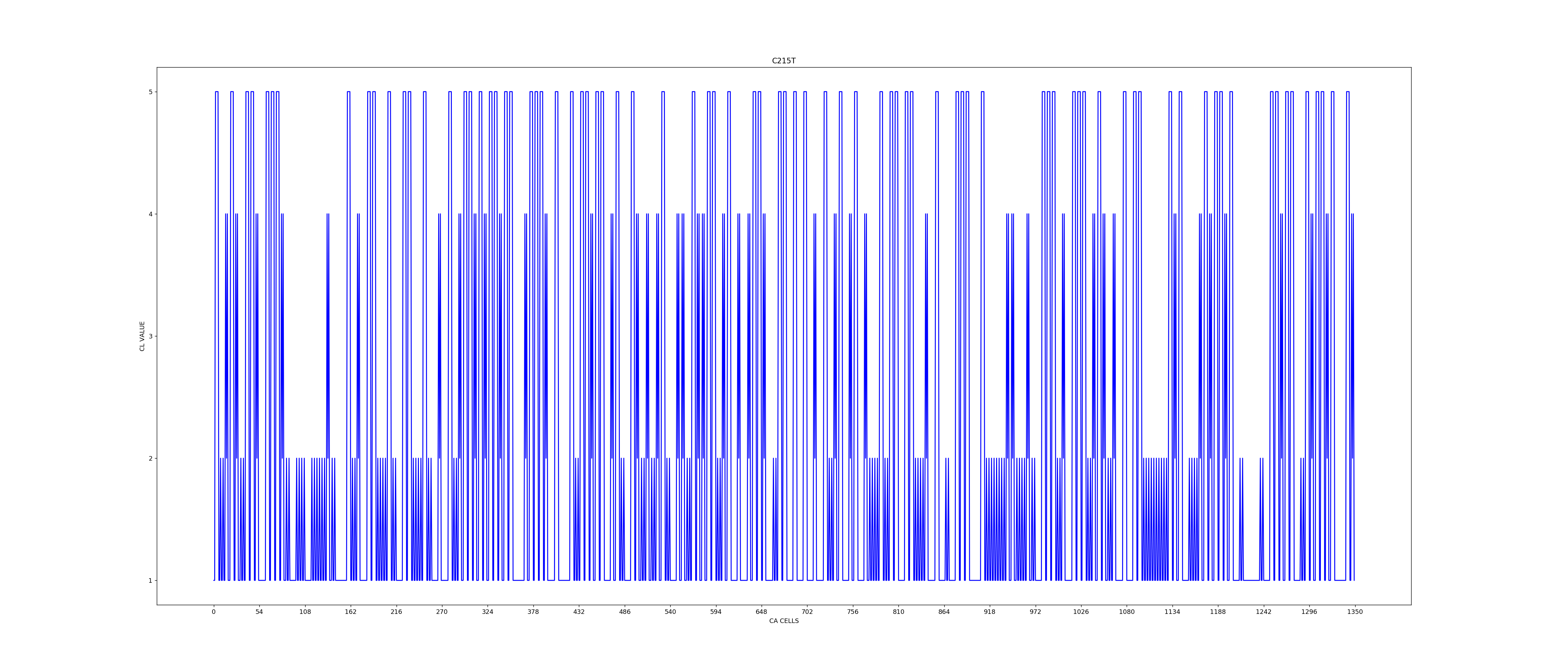}
\caption{Partial CL signal graph for mutant CDS strand with mutation c26304t for CoV envelope protein chain (corresponds to P72L at AA level)}
\label{CL_Graph_Cov_CDS_Mutant_c2640t}
\end{center}
\end{figure}

\begin{figure}[h!]
\begin{center}
\includegraphics[width=13.3cm, height=4.3cm]{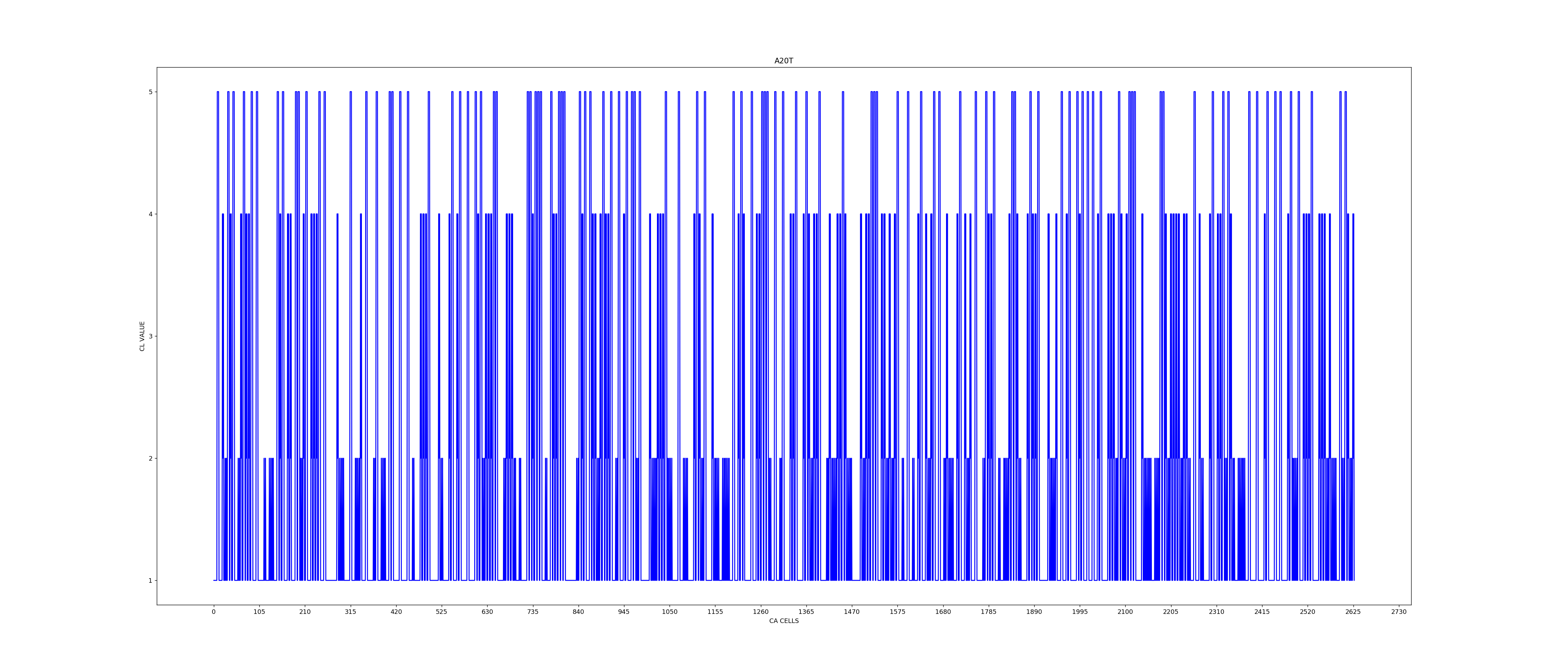}
\caption{Partial CL signal graph for mutant CDS strand with mutation a20t for HBB Hemoglobin protein chain (corresponds to E7V at AA level)}
\label{CL_Graph_HBB_CDS_Mutant_a20t}
\end{center}
\end{figure}

The partial CL Signal graphs derived on evolution of three mutants are shown in Figure~\ref{CL_Graph_Cov2_CDS_Mutant_c2640t}, \ref{CL_Graph_Cov_CDS_Mutant_c2640t} and \ref{CL_Graph_HBB_CDS_Mutant_a20t}, where amino acid Prolin (P) with nucleotide base triplet (cct) mutated with amino acid Leu (L - ctt) for SARS covid, as noted under (a) and (b).

\noindent
(a) mutation P71L for CoV-2 corresponds to c26402t (original nucleotide base c in location 26402 mutated to base t);

\noindent
(b) mutation P72L for CoV corresponds to c26304t (original nucleotide base c in location 26304 mutated to base t);

\noindent
(c) mutation E7V of HBB Hemoglobin has amino acid Glu (E) at serial location 7 corresponds to nucleotide base triplet (gag - at serial base location 19 to 21), while amino acid Val (V) refers to the base triplet (gtg). Hence E7V mutation at AA level corresponds to a20t mutation on CDS strand.

\subsubsection{Evaluation of DCLVS-CDS}

\vspace{2mm}
Similar to evaluation of DCLVS-AA, the difference of CL value sum for CDS strand (DCLVS-CDS) is computed as the difference in CLVS for wild and mutant.

The difference in CL signal graph between wild and mutant of AA chain and CDS strand are marked as parameters DCLVS-AA, DCLVS-CDS. Correspondence of these model parameters to the deviation in structure-function of mutant from the wild type is undertaken through the design of a Machine Learning (ML) framework reported next.

\section{Machine Learning (ML) - A Generic Framework for Prediction of Mutants in a Protein}
\label{Sec:ML_Framework}

Machine Learning (ML) framework has been designed from the analysis of -

\noindent
(a) in vitro/in vivo mutational study reported in published literature, displaying deviation of structure-function of mutant from that of wild for different virus types; and 

\noindent
(b) the parameters DCLVS-AA and DCLVS-CDS derived out of CAML model for these proteins.

\noindent  
This framework enables us to learn the relationship of two model parameters DCLVS-AA and DCLVS-CDS with the deviation of mutant structure/function from its wild version reported in (a).

\subsection{ML Framework Design Methodology}

Ten properties relevant for the framework are highlighted prior to reporting the algorithmic steps of the design.

\subsubsection{The Properties $(P0 ~to~ P9)$}

\vspace{2mm}
These ten properties highlight the underlying principle followed in the ML framework design. 

\vspace{2mm}
\noindent
\textbf{P0} : The threshold values Thv-CDS and Thv-AA are the threshold range for DCLVS-AA reported under P9. A candidate mutant is accepted as registered mutant marked as \emph{RegMut-AA} and \emph{RegMut-CDS} if the threshold range is satisfied. These threshold limits are set from the analysis of model parameters for the mutants reported for vivo/in vitro / in silico mutational studies.  

\vspace{2mm}
\noindent
\textbf{P1} : Similar to the concept of SNP \cite{hancock2008detecting, vitti2013detecting, voight2006map}, only single point mutation is allowed for a non-synonymous mutant.

\vspace{2mm}
\noindent
\textbf{P2} : The CA model employs periodic boundary CA for a protein, considering leftmost amino acid as the neighbour of the rightmost amino acid of the protein chain and vice versa. We concentrate on the analysis of CL signal graph generated out of CA evolution for a protein. The N and C terminal domains of a protein chain are represented on the left and right end of CL signal graph characterized by its mcl and Nmcl signals. 

\vspace{2mm}
\noindent
\textbf{P3} : The CA model, as noted in Section~\ref{CA_rule_forAA}, evaluates backbone and side chain with H and non-H atom count. Based on the atomic structure, the ML framework divides amino acids into two groups - non-ring and ring. The side chain of amino acids Phe (F), Tyr (Y), Trp (W) (Figure~\ref{20_AAcide}) display ring; in addition, Proline (P) is also viewed as a ring since its side chain makes ring with backbone.
  
Mutation from a ring to non-ring amino acid and vice versa is considered as a special mutation class in the ML framework.

\vspace{2mm}
\noindent
\textbf{P4} : Since the chemical properties of amino acids are not considered in the model, the ML framework assigns special consideration for mutation of polar charged amino acid to a non-polar amino acid.

In subsequent discussions, \emph{xLy} refers to a mutant where original amino acid x at location L of amino acid chain gets mutated with amino acid y.

\vspace{2mm}
\noindent
\textbf{P5 (Filter valid mutant)} : The amino acid Gly (G) stands apart from all others with its simplest side chain (one H atom) modelled with simplest CA rule <1>. A mutant xLG is referred to as the conjugate pair of xLy. The Conjugate of a mutant is employed to filter out Valid Mutant (marked in Step 5) out of registered mutants identified in Step 4 of the algorithm reported in Section~\ref{algo}.

\vspace{2mm}
\noindent
\textbf{P6 (Filter valid Mutant as illustrated for HBB protein) :} For two identical mutations xLy, and x(L+1)y on a pair of adjacent amino acids, one of the mutants is a valid one if its Conjugate xLG is a valid mutant. Such filtering properties are employed to match the mutants identified by the CAML model with the mutants reported from in vitro/in vivo mutational study.

\vspace{2mm}
\noindent
\textbf{P7} : In Step 1 of the algorithm, the design sets a flag FF2 (Flag F2) for a mutation xLy, where x and y satisfies one of the following four conditions: (c1) original AA x at location L mutated with y as Met (M); (c2) one of the pair x and y is a ring, while other is a non-ring; (c3) one is a polar charged, while other is an non-polar; and (c4) for the CL signal graphs of wild mutant pair, one has CL(-5), and no CL(-5) for other.

\vspace{2mm}
\noindent
\textbf{P8(a)} (\textbf{Prediction of Mutational Hotspots with CAML model}) : From study of published literature, the list of mutations reported in vitro, in vivo mutational study of a viral protein is compiled and analysed based on - protein length, Nmcl (next to mcl) signal count of CL signal graphs of proteins. Analysis of this experimental data enable CAML platform to set the allowable range of DCLVS-AA and DCLVS-CDS for selection of candidate mutants to be evaluated as possible mutational hotspot in a virus protein.

\vspace{2mm}
\noindent
\textbf{P8(b) (Candidate Mutants to be Evaluated)} : Let xLy be such a mutation with x and y belonging to one of the 5 amino acid groups (Figure~\ref{20_AAcide}). In Step 1 of the design, we concentrate on evaluating a candidate mutant with amino acid pair belonging to groups as that of x and y in locations $L^{'}$ (a location in AA string other than L). In addition, we also evaluate candidate mutants, where x and y belong to same group with 5 non-polar AA (G, A, V, L, I) or 5 polar charged AA (S, T, C, N, Q).

\vspace{2mm}
\noindent                                         
\textbf{P9 (The Allowable Range)} : The algorithm identifies whether a candidate mutant is a valid one on satisfying the following allowable range -

\noindent                                                                                                                                                                (i) For SARS/MERS Covid, set the DCLVS = 50, if FF2 =1 as per P7. For SARS/MERS Covid and HBB proteins : DCLVS - AA value in the range 50 to 200 and DCLVS-CDS value greater than or equal to 7.

\subsection{Design Steps for ML Framework to Identify Valid Mutants}
\label{algo}

\vspace{2mm}
The algorithmic steps to design the Machine Learning (ML) framework of CAML is reported below for an input protein chain and its CDS strand.\\

\vspace{-2mm}
\noindent
\rule[5pt]{1.00\textwidth}{1.00pt}
\textbf{Input :} Amino acid (AA) chain of the wild and candidate mutants on satisfying the conditions reported in Property P8.\\
\rule[5pt]{1.00\textwidth}{1.00pt}\\
\textbf{Step 1 :} \emph{Flag setting} - For a candidate mutant (satisfying Property P8(b)), set the flag FF1 (Flag F1) = 1,  and set  FF2 (Flag F2) = 1 if one of the four conditions (c1 to c4) noted under P7 is satisfied. 

\vspace{2mm}
\noindent
\textbf{Step 2 :} Derive CL signal graphs for wild and mutant CDS strand showing FF1 = 1.

\vspace{2mm}
\noindent
\textbf{Step 3 :} \emph{Identifying RegMutCDS} - Note the mutant for which DCLVS (Difference of CLVS for CDS wild and mutant) is greater than or equal to the threshold limit (Thv-CDS) specified under P9. All these mutants are marked as Registered Mutants at CDS level (referred to as RegMutCDS). Let the cardinality of RegMutCDS be z.

\vspace{2mm}
\noindent
\textbf{Step 4 :} \emph{Identifying RegMutAA} -                                                                                                                                                  \\
(a) Derive the CL signal graphs for wild and z number of mutants identified in Step 3 and evaluate DCLVS-AA (difference of CLVS for AA wild and mutant), where CLVS for mutant is higher than that of wild.\\ 
(b) Select the mutants for which (DCLVS-AA) lies within the threshold range specified under P9. Such a mutant is marked as RegMutAA. 

\vspace{2mm}
\noindent
\textbf{Step 5 :} Apply the filtering property (P5, P6), wherever applicable, to filter out a valid mutant out of registered mutants.

\vspace{2mm}
\noindent
\textbf{Step 6 :} Note mcl count for wild and each RegMutAA. A registered mutant is a valid mutant if its mcl count is greater than or equal to that of wild.

\vspace{2mm}
\noindent
\textbf{Step 7 :} Stop. \\
\rule[5pt]{1.00\textwidth}{1.00pt}

The CAML model identifying ValMut (or simply Mutant) is validated from mutational study results reported in published literature for two case studies.

\newpage
\section{CA Model Validation : Two Case Studies}
\label{Three_Case_studies}

For each of the two case studies, the mutants are identified with the ML framework satisfying the range of threshold values Thv-CDS and ThV-AA specified under P9. This list of mutants covers the mutants reported in vitro/in vivo mutational studies, where structure-function of the mutant differ from that of wild. A mutation xLy (amino acid x at location L mutated to y) corresponds to the CDS mutation $x^{'}L^{'}y^{'}$, where $L^{'} = (3L-2)$ or $(3L-1)$ or $(3L)$ , and $x^{'}$,  $y^{'}$ are nucleotide bases derived out of amino acid x and y respectively. Single point mutation is identified for the CDS strand that corresponds to mutation xLy at amino acid level.

\subsection{Case Study 1 on Hemoglobin HBB beta-globin protein \cite{cai2018universal, NatureEducation2008, ncbiDatabase, carlice2017investigation}}

\begin{figure}[h!]
\begin{center}
\includegraphics[width=13.3cm, height=4.9cm]{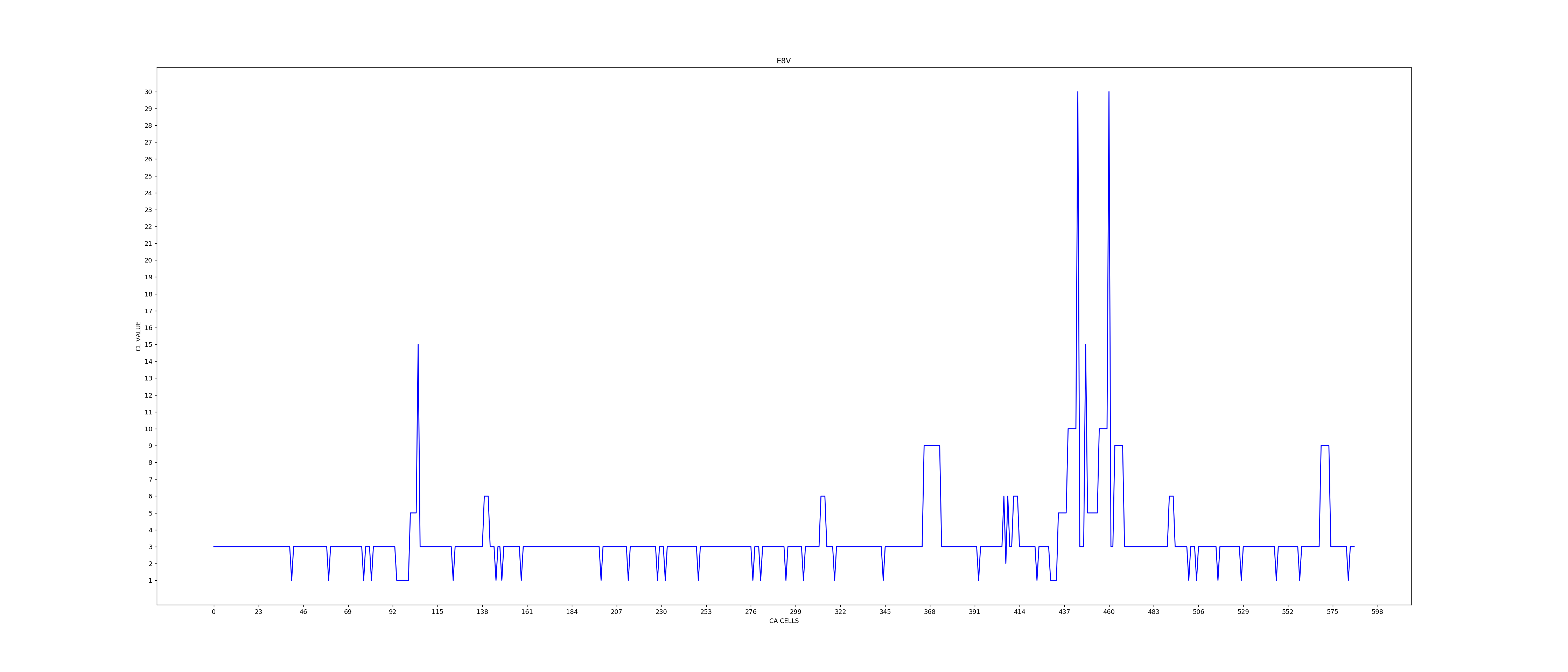}
\caption{HBB Mutant E8V - AA level with mcl count 2}
\label{CL_Graph_HBB_AA_Mutant_E8V}
\end{center}
\end{figure}

\begin{figure}[h!]
\begin{center}
\includegraphics[width=13.3cm, height=4.1cm]{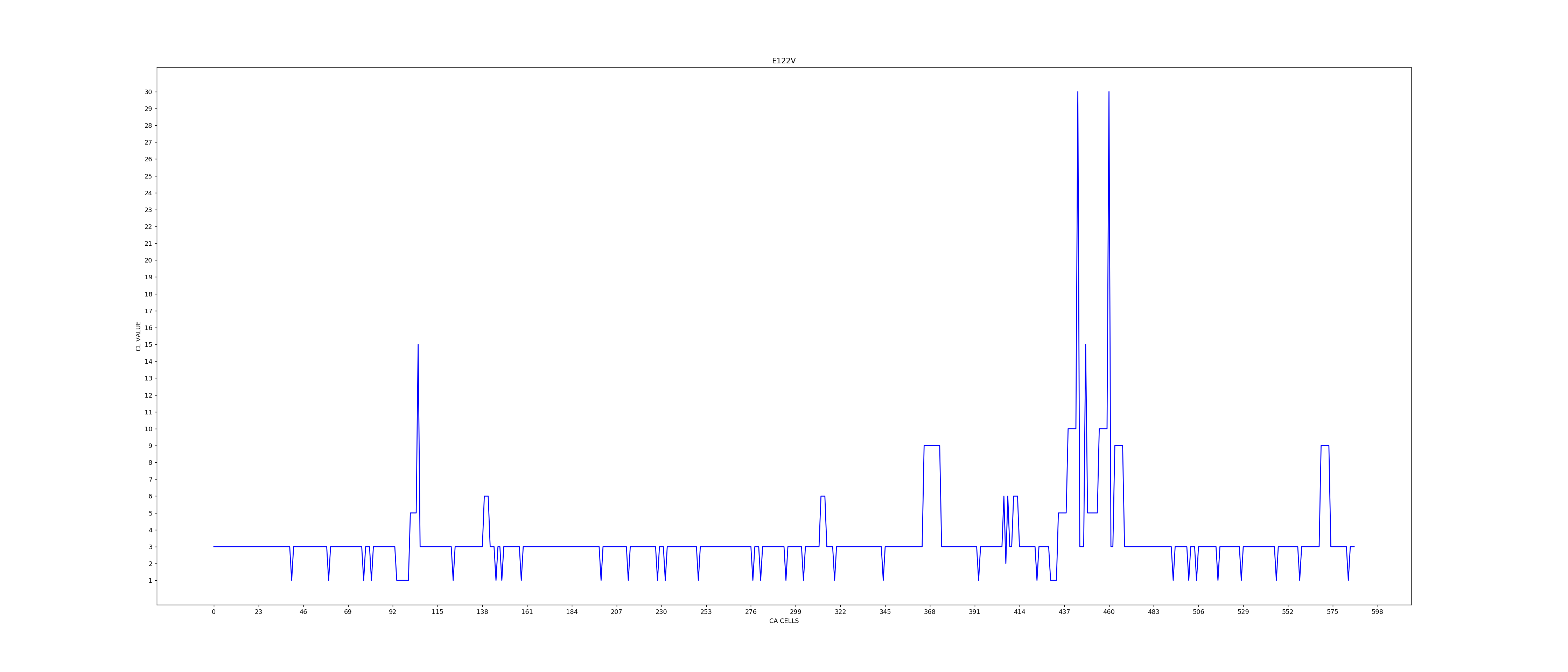}
\caption{HBB Mutant E122V - AA level with mcl count 2}
\label{CL_Graph_HBB_AA_Mutant_E122V}
\end{center}
\end{figure}

\begin{figure}[h!]
\begin{center}
\includegraphics[width=13.3cm, height=4.1cm]{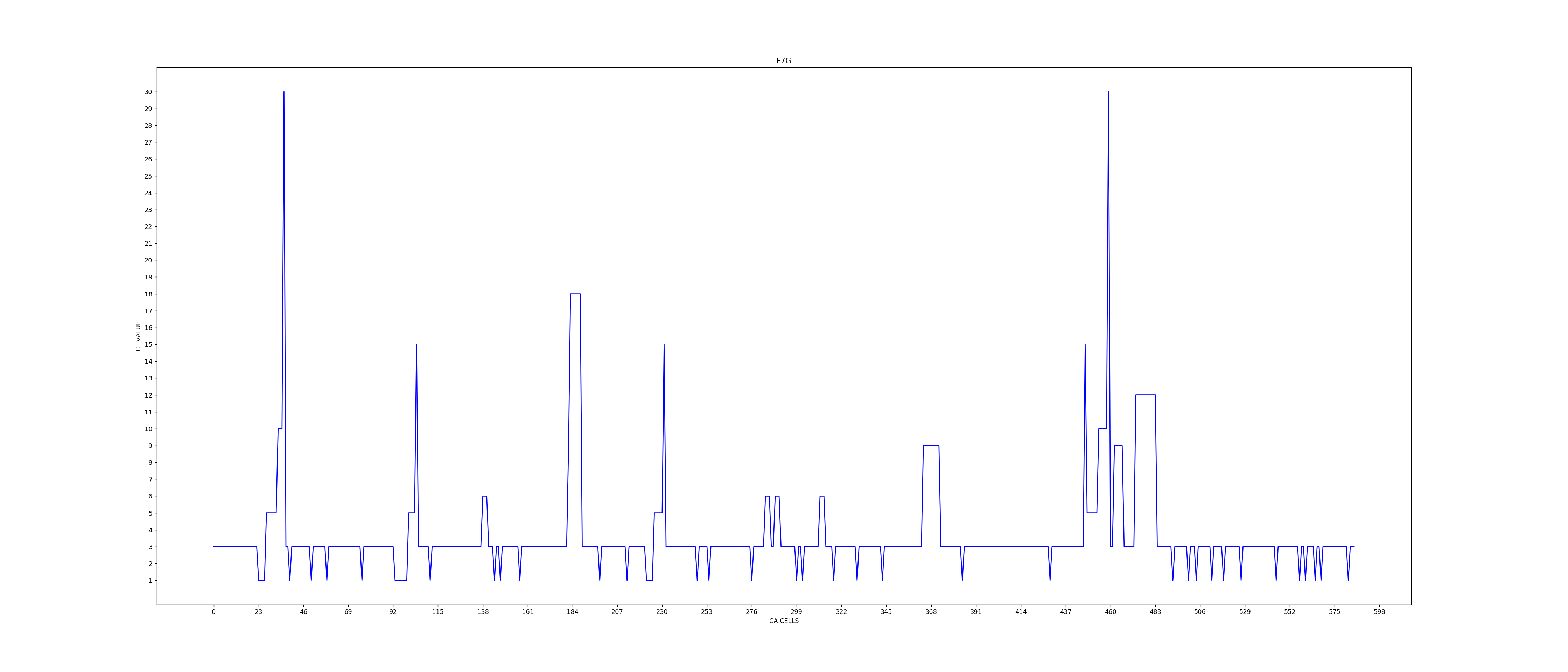}
\caption{HBB Mutant E7G - AA level with mcl count 2}
\label{CL_Graph_HBB_AA_Mutant_E7G}
\end{center}
\end{figure}

\begin{figure}[h!]
\begin{center}
\includegraphics[width=13.3cm, height=4.1cm]{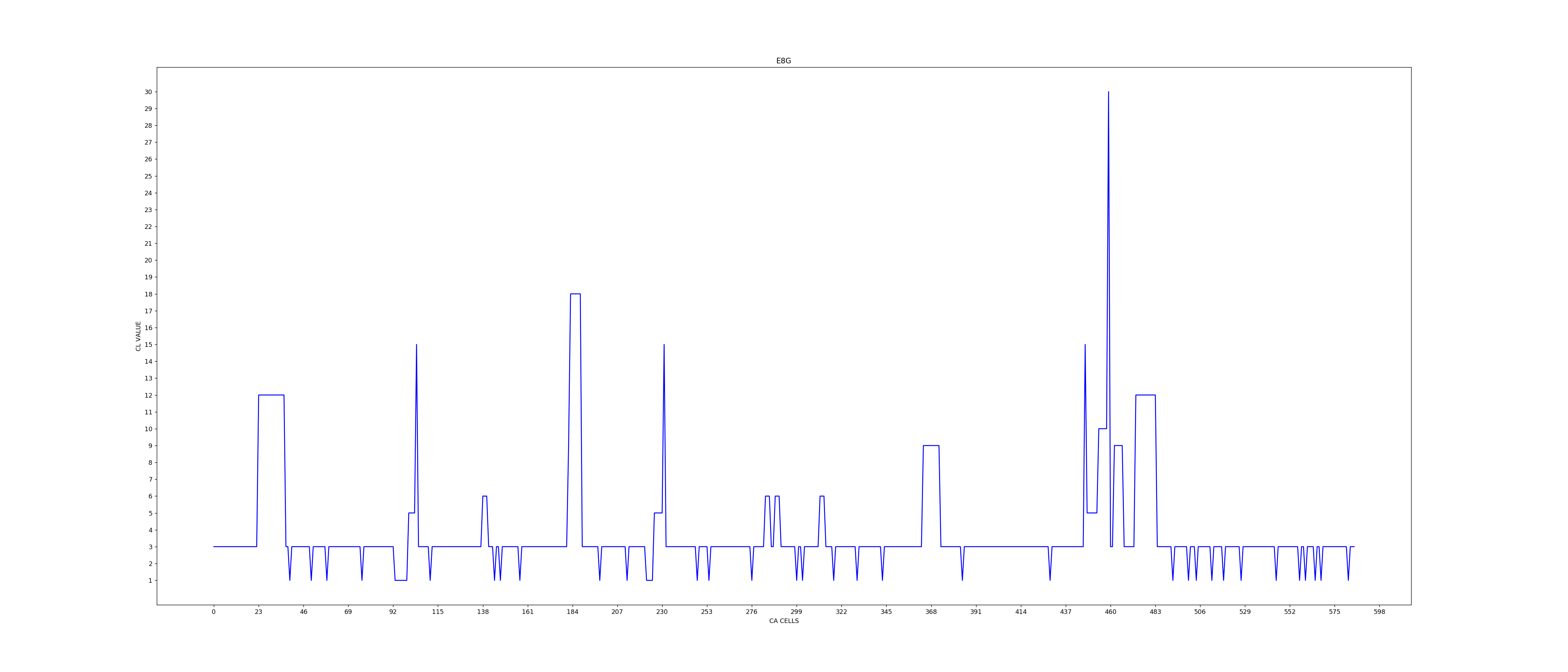}
\caption{HBB Mutant E8G - AA level with mcl count 1}
\label{CL_Graph_HBB_AA_Mutant_E8G}
\end{center}
\end{figure}

\vspace{2mm}
This study identifies two valid mutants E7V, E122V with polar negatively charged amino acid glutamate (E) replaced with nonpolar valine (V) and so FF2 is set to 1 in Step 1 of the algorithm. Exhaustive study of E7V mutant has been reported in many publications \cite{NatureEducation2008, carlice2017investigation}. Figure~\ref{CL_Graph_Wild_AA_HBB}  shows CL Signal graph at AA level for wild with mcl count 2, and Figure~\ref{CL_Graph_HBB_AA_Mutant_E7V}, \ref{CL_Graph_HBB_AA_Mutant_E8V} and \ref{CL_Graph_HBB_AA_Mutant_E122V} for three mutants E7V (FF2 = 1, DCLVS-CDS = 34 (corresponds to a20t)), E8V (FF2 = 1, DCLVS-CDS = 34 (corresponds to a23t), E122V (FF2 = 1, DCLVS-CDS = 14 (corresponds to a365t) - these 3 mutants are marked as RegMutAA. Figure~\ref{CL_Graph_HBB_AA_Mutant_E7G} and \ref{CL_Graph_HBB_AA_Mutant_E8G} shows two mutants E7G with mcl count 2, and E8G with mcl count 1. As per Step 6, the mutant E8G is discarded - it is not a valid mutant since its mcl count 1 (see Figure~\ref{CL_Graph_HBB_AA_Mutant_E8G}) is less than wild mcl count 2 (see Figure~\ref{CL_Graph_Wild_AA_HBB}). Figure~\ref{CL_Graph_Wild_CDS_HBB} and \ref{CL_Graph_HBB_CDS_Mutant_a20t} display HBB CL graph at CDS level for wild and HBB mutant a20t showing DCLVS-CDS = 34 (3+4+9+9+9 at the region covering cell locations 150 to 154 of CL signal graph of CDS strand). In addition to E8V, mutations of E with V at AA location 23, 27, 44, 91, 102 of HBB amino acid string (of length 147) are also discarded since mcl count for Gly mutation (Property P5) for each of the E at these locations show mcl cnt 1.

In addition to E7V and E122V, the ML frame work also identifies the following two as valid mutants - (i) E27K (with DCLVS-AA = 489 - 361 = 124, and DCLVS-CDS = 7 (corresponds to a80c)), and (ii) P6S (FF2 = 1, and DCLVS-CDS = 9 (corresponds to c16t)). These two mutations are reported in the in silico mutational study \cite{carlice2017investigation}. However, the following 5 mutations are also reported in this study which are not detected as valid mutant by the ML framework - E7K, R31S, N109H, E122Q, A130V. The number of patients for whom such mutants were observed are low (Table 1 in \cite{carlice2017investigation}). Detailed investigation of mutational study on HBB Hemoglobin beta-globin protein is beyond the scope of the current paper.

\subsection{Case Study 2 on CoV-2 Envelope Protein \cite{petersen2020comparing, hassan2020sars, ncbiDatabase}}

\begin{figure}[h!]
\begin{center}
\includegraphics[width=13.3cm, height=4.4cm]{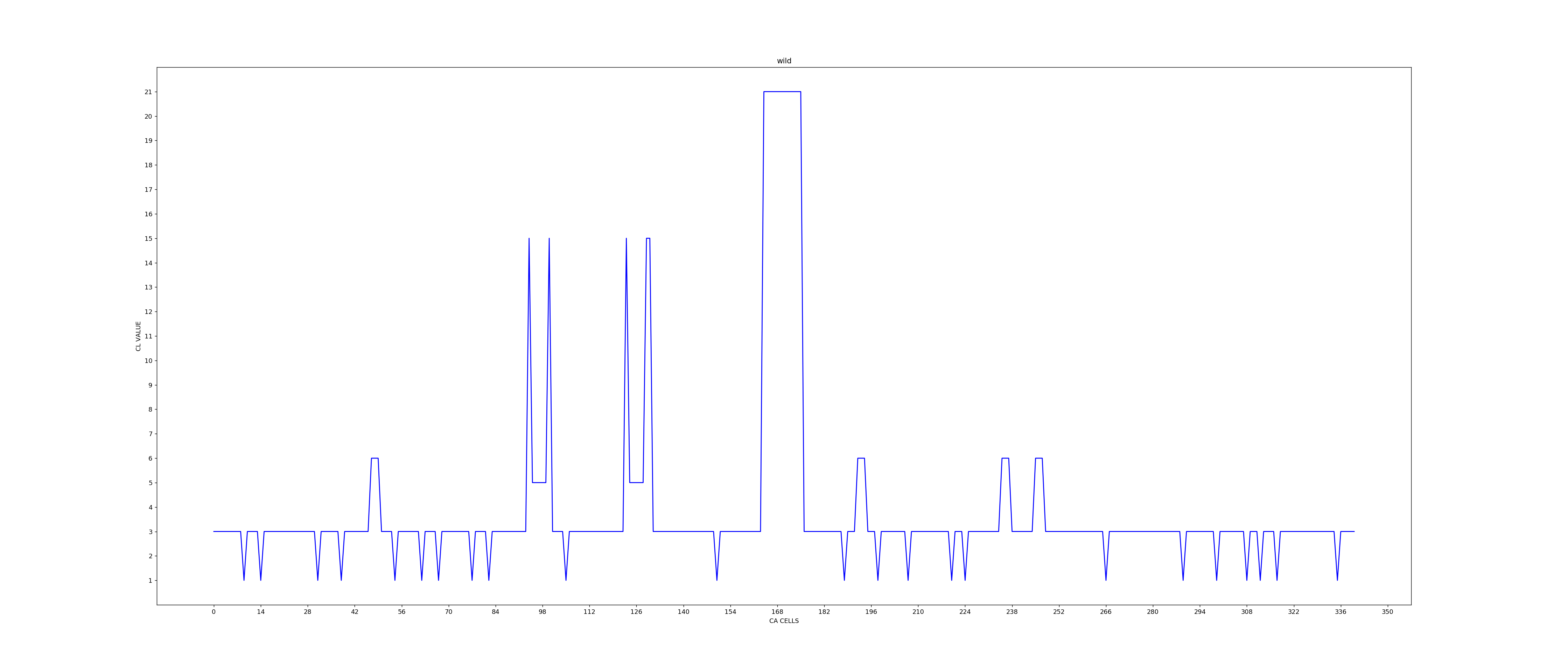}
\caption{CL signal graph for wild envelope protein of MERS AA chain}
\label{CL_Graph_Wild_AA_MERS}
\end{center}
\end{figure}

\begin{figure}[h!]
\begin{center}
\includegraphics[width=13.3cm, height=4.4cm]{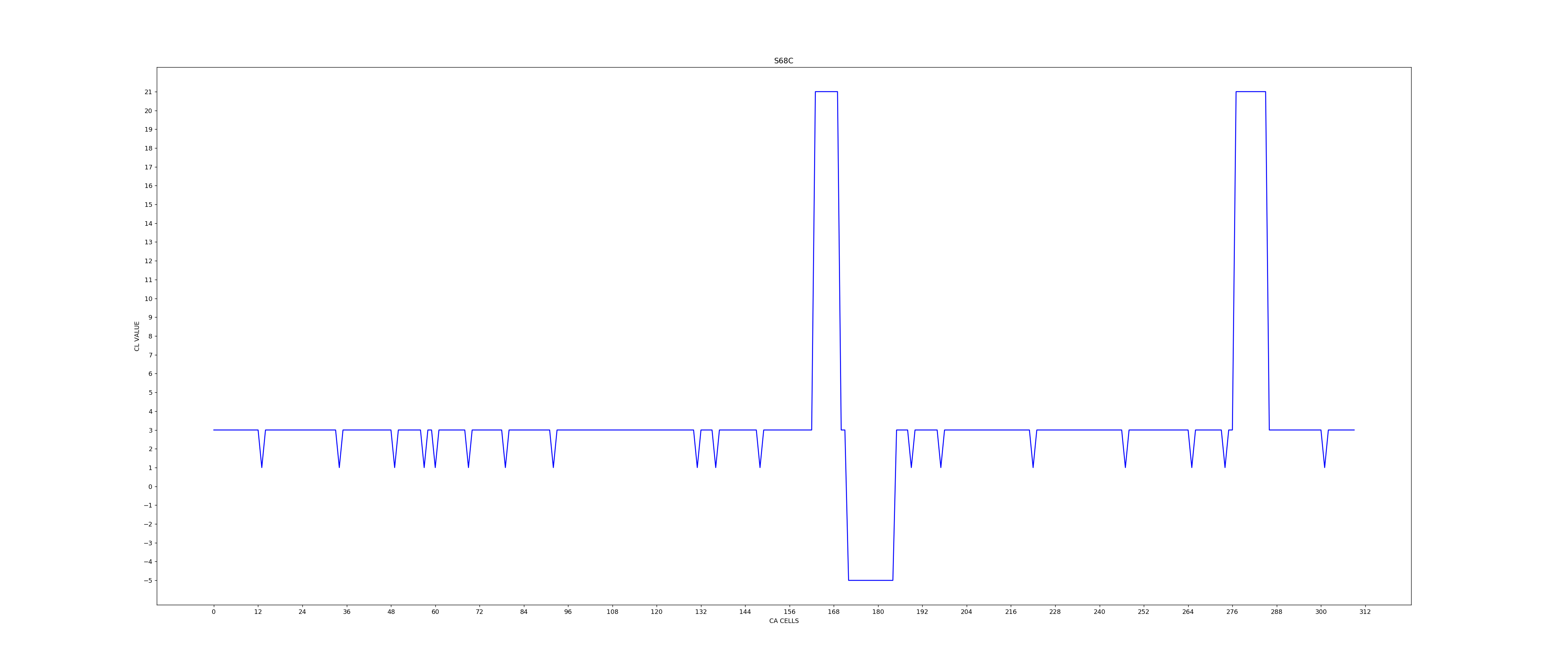}
\caption{CoV-2 mutant S68C : DCLVS with wild = 9 x 18 =  162; signal value 18 (width 9 cell) at loc 278 to 287}
\label{CL_Graph_Cov2_AA_Mutant_S68C}
\end{center}
\end{figure}

\begin{figure}[h!]
\begin{center}
\includegraphics[width=13.3cm, height=4.4cm]{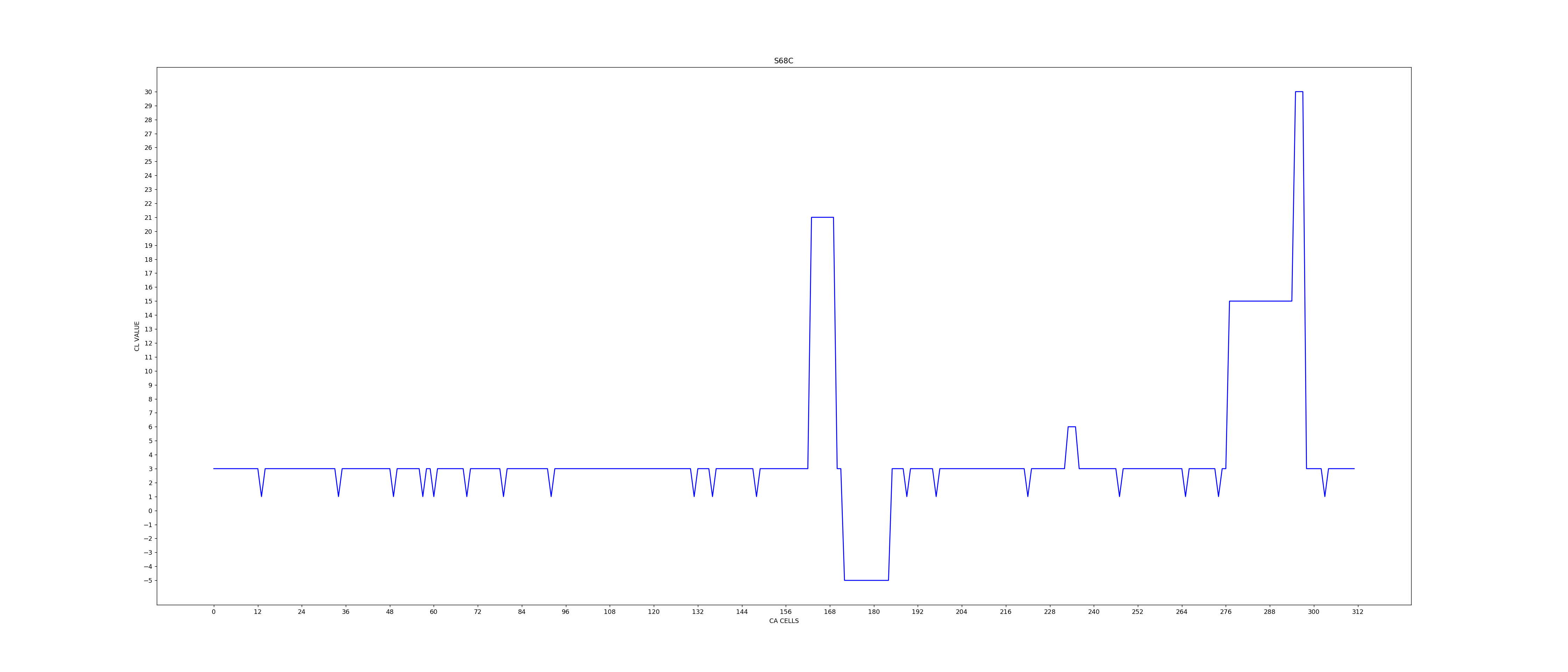}
\caption{CoV mutant S68C : DCLVS-AA = 18 x 15 + 3 x 30 = 297; signal at cell location 276 to 297 with CL value 15 (width 18 cells at cell loc 276 to 294) and CL value 30 (width 30 at cell loc 294 to 297) }
\label{CL_Graph_Cov_AA_Mutant_S68C}
\end{center}
\end{figure}

\vspace{2mm}
SARS-CoV-2 (2019) wild envelope Protein with 75 amino acid chain reported in Table~\ref{AA_Chain} . Figure~\ref{CL_Graph_Wild_AA_Cov2}, \ref{CL_Graph_Wild_AA_Cov} and \ref{CL_Graph_Wild_AA_MERS} display CL Signal Graph for wild CoV-2, CoV and MERS respectively, and Figure~\ref{CL_Graph_Cov2_AA_Mutant_S68C}, \ref{CL_Graph_Cov_AA_Mutant_S68C} for CoV-2 and CoV mutants S68C. Six mutants (A36V, L37H, L39M, S68F, D72Y, L73F) are also identified. All these 7 (2 + 5) mutants are identified in Envelope Protein for CoV-2 infected patients reported in \cite{lee2010pdz, hassan2020sars}.  Table~\ref{Cov2_7Valid_Mutant} third column displays the parameter DCLVS (Difference of CL Value sum) between wild and each of these 7 mutants. For six of these mutants the flag FF2 = 1, as per the conditions noted under property P7.  Hence DCLVS is set as 50,  as per Property P9(i). Table~\ref{Cov2_additional_mutant} reports six additional mutants predicted with ML framework. For all these mutants the parameters DCLVS-AA lies within the range 50 to 200 and DCLVS-CDS display threshold value greater than or equal to 7 respectively. Mutant S16C is discarded since its DCLVS-AA lies outside the range (50 to 200).

\begin{table}[h!]
\centering
\resizebox{0.9\textwidth}{!}{
\begin{tabular}{|c|c|c|}
\hline 
\multirow{2}{*}{Mutant} & DCLVS & \multirow{2}{*}{Comment} \\  
 & (mutant CLVS - wild CLVS) &  \\ 
\hline 
L73F & 50 & FF2 = 1, sets DCLVS = Thv = 50 (as per P9(i)) \\ 
\hline 
P71L & 50 & FF2 = 1, sets DCLVS = Thv = 50 (as per P9(i)) \\ 
\hline 
D72Y & 50 & FF2 = 1, sets DCLVS = Thv = 50 (as per P9(i)) \\ 
\hline 
S68F & 50 & FF2 = 1, sets DCLVS = Thv= 50 (as per P9(i)) \\ 
\hline 
S68C & 162 & FF2 = 0  \\ 
\hline 
L39M & 50 & FF2 = 1, sets DCLVS = Thv = 50 (as per P9(i)) \\ 
\hline 
A36V & 50 & FF2 = 1, Cl(-5) for wild, not for mutant(as per P8) \\ 
\hline 
\end{tabular} }
\caption{DCLVS difference between wild and mutant CL value sum for 7 mutants are the mutants identified by ML framework}
\label{Cov2_7Valid_Mutant}
\end{table}

\begin{table}[h!]
\centering
\resizebox{0.85\textwidth}{!}{
\begin{tabular}{|c|c|c|}
\hline 
\multirow{2}{*}{Mutant} & DCLVS & \multirow{2}{*}{Comment} \\  
 & (mutant CLVS - wild CLVS) &  \\ 
\hline 
L26F & 50 & instance of FF2 = 1, sets DCLVS = Thv = 50 (as per P9(i)) \\ 
\hline 
S16C & 231 & FF2 = 0, invalid since DCLVS outside the range 50 to 200 \\ 
\hline 
V58G & 132 & FF2 = 0 \\ 
\hline 
N64Y & 50 & FF2 = 1, sets DCLVS = Thv = 50 (as per P9(i)) \\ 
\hline 
S60P & 50 & FF2 = 1, sets DCLVS = Thv = 50 (as per P9(i)) \\ 
\hline 
S60Y & 50 & FF2 = 1, sets DCLVS = Thv = 50 (as per P9(i)) \\ 
\hline 
C40H & 50 & FF2 = 1, sets DCLVS = Thv = 50 (as per P9(i)) \\ 
\hline 
L73C & - & not a mutant - needs two mutations at CDS level  \\ 
\hline 
A22C & - & not a mutant - needs two mutation at CDS level  \\ 
\hline 
A22D & 0 & not a mutant - DCLVS-AA outside the range 50 to 200 \\ 
\hline 
\end{tabular} } 
\caption{Six mutants (first seven rows other than 2nd row mutant S16C) are identified as (valid) mutants as per ML framework, last 3 mutants are invalid due to DCLVS less than threshold value 50 or it needs two point mutations at CDS level}
\label{Cov2_additional_mutant}
\end{table}

\section{Comparison of Transmissibility of SARS Cov-2, SARS CoV, and MERS Covid -- Based on CAML Model of Envelope Proteins}
\label{Comparison}

Deviation of structure - function of a mutant from its wild version gets reflected in the high value (greater than or equal to 50) of CAML model parameter DCLVS-AA. The same framework is also relevant for evaluation of DCLVS for a pair of proteins executing similar function and length ( difference less than 10 \%). First three rows of Table~\ref{Comparison_of_wild_pairs} compare the DCLVS-AA of a pair of wild envelop proteins of (CoV-2, CoV, MERS). Large difference (90) in CLVS value between CoV-2 wild and CoV/MERS wild confirms large difference in transmissibility of CoV-2 compared to CoV and MERS. On the other hand, difference in CLVS between CoV and MERS (third row of Table~\ref{Comparison_of_wild_pairs}) is nil and so transmissibility of these two envelop proteins are of the same order. Two split Nmcl signals in cell locations (93 to 98) and (121 to 129) of MERS wild CL graph (Figure~\ref{CL_Graph_Wild_AA_MERS}) are excluded, as per Section~\ref{Evaluation_of_Difference}, to compare CLVS of MERS with CoV and CoV-2.

\begin{table}[h!]
\centering
\resizebox{1.0\textwidth}{!}{
\renewcommand{\arraystretch}{1.1}
\begin{tabular}{|c|c|c|}
\hline 
\textbf{Pairs of Envelop proteins} & \textbf{DCLVS} & \textbf{Comments} \\ 
\hline 												
\multirow{4}{*}{Wild CoV-2 \& Wild CoV} & \multirow{4}{*}{$10 \times (12 - 3) = 90$} & difference of CL value 12 (width 10 cells) at    \\ 
& & cell loc 280 to 290; for both CoV-2 and   \\
& & CoV, mcl signal of value 21 (width 7 cells)  \\
& & at cell loc 160 to 167  \\
\hline 
\multirow{4}{*}{Wild CoV-2 \& Wild MERS} & \multirow{4}{*}{$5 \times (21 - 3) = 90$} & CL value difference in the region covering   \\ 
 & &  cells at location 160 to 165 MERS, on exclusion   \\
 & & of 2 Split Nmcl signals, mcl signal of   \\
 & & value 21 (width 12) at cell loc 166 to 178 \\
\hline 
\multirow{3}{*}{Wild CoV \& Wild MERS} & $5 \times (21 - 3) -$ & CL value difference for the cells 160 to 165   \\
 & $10 \times (12 - 3) = 0$ & and 280 to 290 CoV and MERS on   \\
 & & exclusion of 2 Split Nmcl signals  \\ 
\hline
Wild CoV-2 \& CoV mutant & \multirow{2}{*}{0} & same CL graph for CoV-2 and CoV with    \\
with deletion of G at loc 70 & & deletion of Gly (G) at location 70 \\
\hline 
CoV-2 with R at loc 69 & \multirow{2}{*}{0} & same CL graph for mutated CoV-2  \\
replaced with EG \& CoV wild & & and CoV wild  \\
\hline
\end{tabular} }
\caption{\small{Comparison of wild pairs of (CoV-2, CoV), (CoV-2, MERS), (CoV, MERS) on first three rows. Fourth row confirms that mutation of CoV with deletion of G at location 70 displays same CL graph as that of CoV-2. Fifth row confirms that CoV wild CL graph is identical  with mutated  CoV-2 on replacing amino acid R at location 69 with amino acid pair EG at location at 69-70.}}
\label{Comparison_of_wild_pairs}
\end{table}

\subsection{Analysis of signals in the C-terminal domain of virus covering Nmcl signal value 12 at the region covering cell locations 280 to 290 of CL signal graph of CoV wild} 

\vspace{2mm}
The C-terminal domain of virus plays crucial role in virus life cycle. The C-terminal domain of SARS CoV-2 (AA loc 39 to 75, cell loc 158 to 300) and CoV (39 to 76, cell loc 158 to 304) covering mcl signal and Nmcl (next to mcl) signal.  On the other hand MERS (82 AA) C-terminal domain covers the cell locations 153 to 328.

The CL signal graphs of CoV-2, CoV, and MERS wild are displayed in Figure~\ref{CL_Graph_Wild_AA_Cov2}, \ref{CL_Graph_Wild_AA_Cov} and \ref{CL_Graph_Wild_AA_MERS}. The Nmcl signal of value 12 at the region covering cell locations 280 to 290 of CoV is a determining factor for comparison of DCLVS between CoV-2 and CoV. This region covering cell locations 280 to 290 is marked as  Virus - Host Golgi Complex - Interaction Site (V-HGC-IS) in the C-Terminal domain of virus CL graph. Absence of the Nmcl signal at V-HGC-IS for CoV-2 indicates stronger interaction of CoV-2 with host Golgi complex resulting in packing/assembly of virus with higher efficiency. On the other hand, presence of Nmcl signal at V-HGC-IS confirms  weaker interaction of CoV with Golgi complex resulting in packaging/assembly of virus with efficiency lesser than that of CoV-2. This difference in interaction of CoV-2 and CoV with host Golgi complex points to higher transmissibility of CoV-2 compared to CoV.

So far as virus-host interaction is concerned, MERS stands apart from  CoV-2 and CoV \cite{hu2020comparison, zhu2020sars}. The MERS wild CL graph does not show any signal in the region covering cell location 280 to 290. However, MERS mcl signal width differs from that of CoV-2 and CoV in the region covering cell locations 166 to 178 in its C-terminal domain spanning from cell location 153 to 328. The C-terminal domain of SARS covid starts from cell location 158 and spans up to cell location 300/304 with  mcl signal (width 7 cells) covered by cell locations 160 to 167. As a result, DCLVS between CoV and MERS covering C-terminal domain, as shown on third row of Table~\ref{Comparison_of_wild_pairs} is nil. This result confirms same level transmissibility for CoV and MERS.

\subsection{Role of amino acid Glu (E ) and Gly (G) at location 69-70 of SARS CoV and amino acid R at location 69 of CoV-2 : the determining factor for higher transmissibility of CoV-2 compared to CoV}

\label{Determining_factor}

\begin{figure}[h!]
\begin{center}
\includegraphics[width=13.3cm, height=4.9cm]{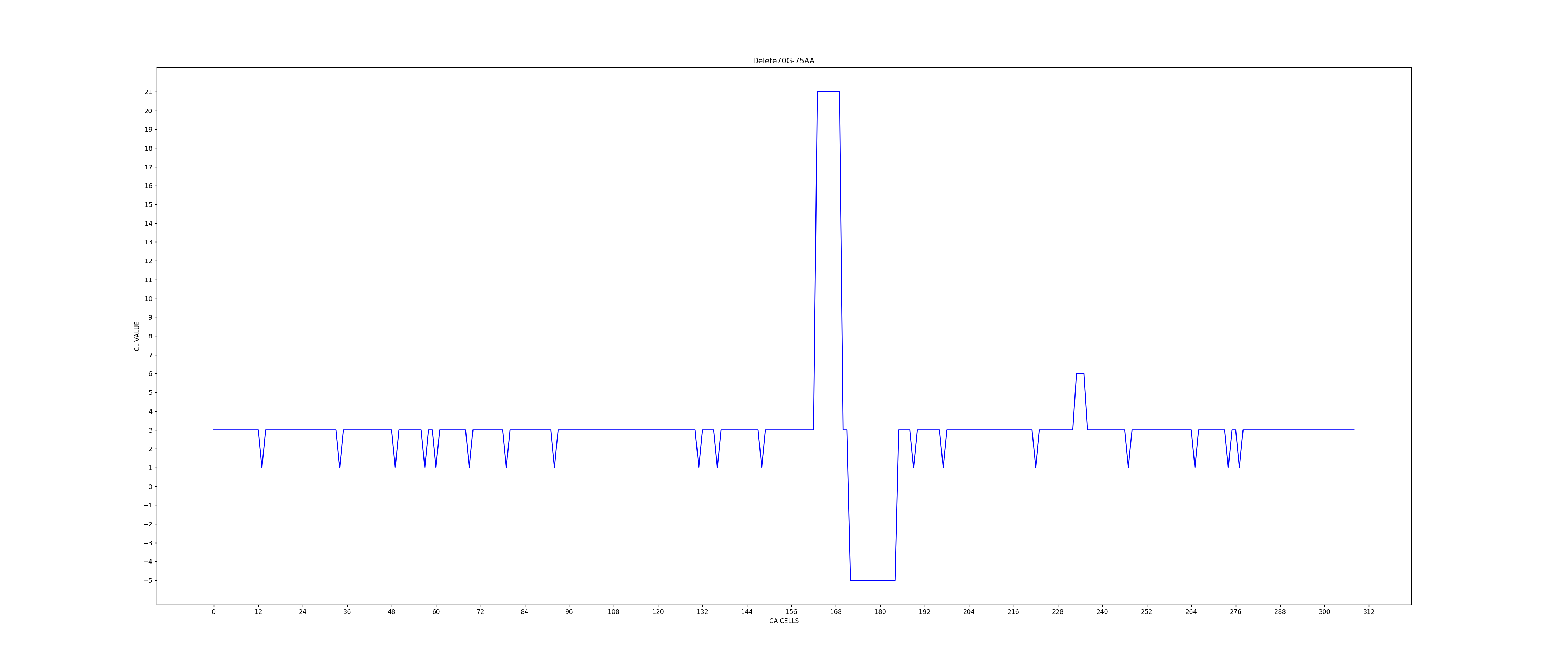}
\caption{CoV(2003) Envelope Protein CL Signal Graph on deleting amino acid G at location 70 - makes its length 75 (Identical to CL graph for CoV-2 (see Figure~\ref{CL_Graph_Wild_AA_Cov2}))}
\label{CL_Graph_Cov_AA_Delete70G}
\end{center}
\end{figure}

\begin{figure}[h!]
\begin{center}
\includegraphics[width=13.3cm, height=4.9cm]{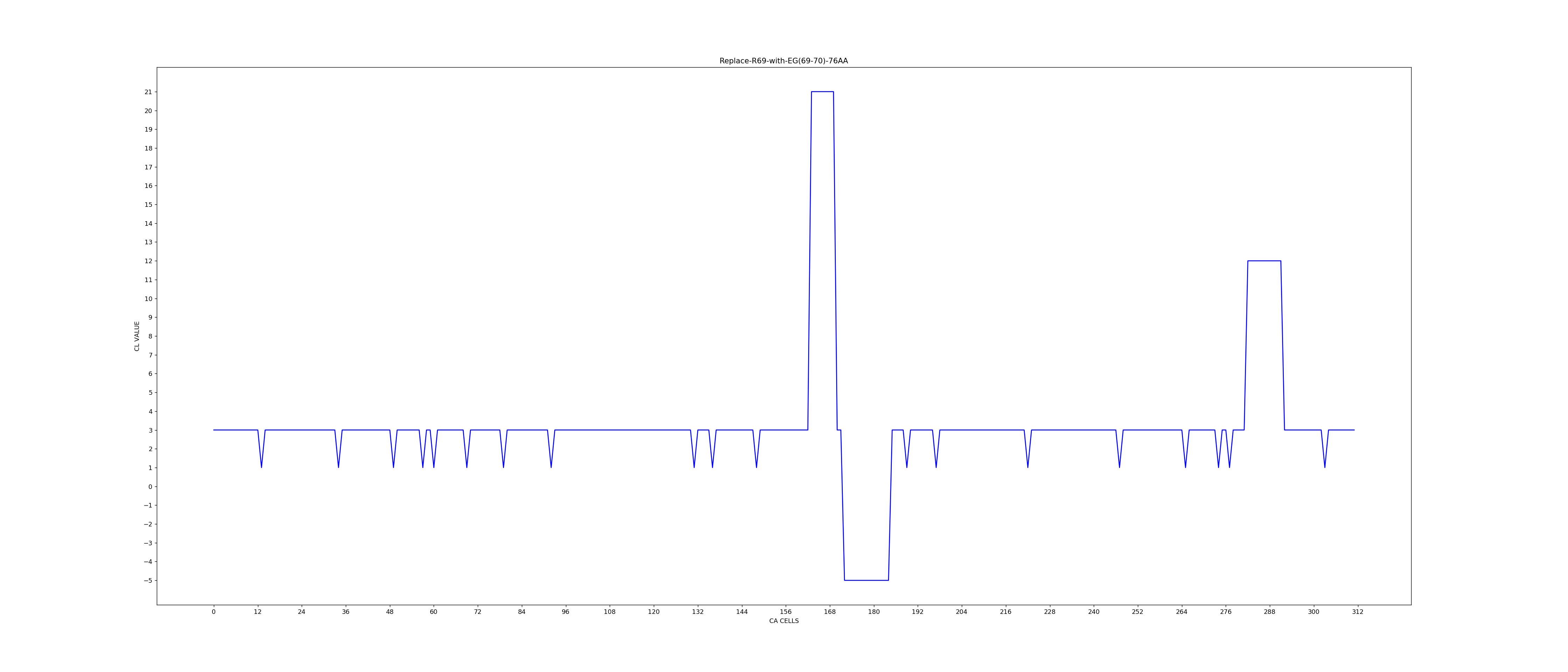}
\caption{CL graph for mutated CoV-2 with R at location 69 replaced with amino acid pair EG at location 69 - 70 making length 76; this CL graph of mutated CoV-2 is identical to CoV wild CL graph (see Figure~\ref{CL_Graph_Wild_AA_Cov})}
\label{CL_Graph_Cov2_AA_R69_with_EG}
\end{center}
\end{figure}

\vspace{2mm}
CoV-2 and CoV Envelope Proteins differ in loc 55, 56 (CoV-2 has S, F in place of T, V for CoV), and location 69, 70 (CoV-2 has R at location 69, while CoV shows E, G). Deletion of amino acid G at location 70 for CoV, makes its length 75. Figure~\ref{CL_Graph_Cov_AA_Delete70G} shows CL graph for CoV on deleting the amino acid G. This CoV mutant (with deleted G) shows large difference of CL signal graph with its wild (Figure~\ref{CL_Graph_Wild_AA_Cov}) in the region covering cell locations 280 to 290. As reported on row 4 of Table~\ref{Comparison_of_wild_pairs}, this mutant has nil difference with CoV-2 wild (Figure~\ref{CL_Graph_Wild_AA_Cov2}) CL graph. The row 5 of Table~\ref{Comparison_of_wild_pairs} compares CoV-2 mutant (with amino acid R replaced with amino acid pair EG) and wild CoV, both having length 76. The Figure~\ref{CL_Graph_Cov2_AA_R69_with_EG} displays the CL signal graph of CoV-2 mutant with amino acid R at location 69  replaced with amino acid pair EG at locations 69-70, making its length 76. This CL graph of CoV-2 mutant match  with that of CoV. These results establish the crucial role of amino acid pair EG at location 69-70 of CoV envelope protein. Replacement of this amino acid pair EG with amino acid R at location 69 of CoV-2 plays as a determining factor for high transmissibility of CoV-2 compared to CoV.

\subsection{Comparison of DCLVS for Mutations inserted in CoV-2 and CoV}

\vspace{2mm}
Table~\ref{Identified_valid_mutant} rows 3 to 8 report six mutants identified by ML framework for CoV-2 and CoV with mutations inserted in identical amino acid type on same or adjacent location. The CAML parameters DCLVS-AA and DCLVS-CDS for each of these mutants satisfies the threshold limit 50 and 7 respectively and so these are valid mutants. However, mutations noted on first two rows of Table~\ref{Identified_valid_mutant} are not valid mutants since one of the parameters DCLVS-AA  or  DCLVS-CDS value is less than the specified threshold value. Further, CoV-2/CoV mutants D73F/D74F (last row of Table~\ref{Identified_valid_mutant}) demands double mutations and so discarded. The CL signal graphs for mutant S68C for CoV-2 and CoV are shown in Figure~\ref{CL_Graph_Cov2_AA_Mutant_S68C} and Figure~\ref{CL_Graph_Cov_AA_Mutant_S68C}. The signal at the region covering cell locations 278 to 298 of CoV is due to amino acid pair EG (discussed in Section~\ref{Determining_factor}) at location 69-70 of CoV envelop protein chain. The large DCLVS value (90 and 135) for each of the valid 6 mutants for CoV-2 and CoV confirm large deviation of structure - function of CoV-2 from CoV that points to large difference in transmissibility of CoV-2 compared to CoV. 

High DCLVS value between wild SARS CoV-2 and SARS CoV/MERS covid chains and also for their mutants in identical amino acid type establishes large deviation of structure - function of CoV-2 (2019) from that of CoV (2003). The deviation of structure - function gets reflected in the large difference in their transmissibility.

\begin{table}[h!]
\centering
\resizebox{1.0\textwidth}{!}{
\renewcommand{\arraystretch}{1.1}
\begin{tabular}{|c|c|}
\hline 
\textbf{Mutants} & \multirow{2}{*}{DCLVS} \\  
\textbf{CoV-2/Cov} &  \\ 
\hline  
L37H/L37H & 90; same difference as of wild CoV-2 and CoV at loc 280 to 290  \\ 
\hline 
L39M/L39M  & $10 \times (12 - 3) = 90$; difference at loc 282 to 291   \\ 
\hline 
\multirow{2}{*}{S68C/S68C} & 297 - 162 = 135; CoV CLVS = $18 \times 15 + 3 \times 30 = 297$ (loc 276 to 297) for Figure~\ref{CL_Graph_Cov_AA_Mutant_S68C}  \\  
 & and CoV-2 CLVS = $9 \times 18 = 162$ (loc 278 to 287) in Figure~\ref{CL_Graph_Cov2_AA_Mutant_S68C} \\ 
\hline 
S68F/S68F & 90; same difference as Wild for CoV-2/CoV at loc 280 to 290 \\ 
\hline 
P71L/P72L & $10 \times (12 - 3) = 90$; same difference as Wild for CoV-2/CoV at loc 280 to 290 \\ 
\hline 
D72Y/D73Y & 90; same difference as Wild for CoV-2/CoV at loc 280 to 290 \\ 
\hline 
D73F/D74F & (not covered by ML framework); D = ga-c/t, F = tt-c/t - need two point mutations \\ 
\hline 
\end{tabular} }
\caption{CoV-2/CoV mutants with mutation inserted on identical or adjacent amino acid locations with same AA type - all the 6 mutants reported from $1^{st}$ to $6^{th}$ row are identified as valid mutants by ML framework satisfying DCLVS-AA threshold value range (+50 to +200) and greater than or equal to 7 for  DCLVS-CDS }
\label{Identified_valid_mutant}
\end{table}

\section{Conclusion}

High transmissibility of CoV-2, as per our analysis, is a combined effect of mutations on structural proteins (Envelope E, Spike S, and Nucleocapsid N) and the role played by non-structural proteins (nsps) and accessory proteins (ORFs). Viral life cycle in host cell is controlled by these proteins for replication and host immune evasion. The current paper is the first one in the series of three papers (we shall report) investigating the reason of highest transmissibility of SARS CoV-2 (2019) compared to SARS CoV(2003) and MERS (2012). The Cellular Automata enhanced Machine Learning (CAML) model is presented in the current paper to investigate the contribution of envelop protein for high transmissibility of SARS CoV-2. The Machine Learning (ML) framework is designed to learn the threshold limit of CA model parameter so that the list of mutants identified by the ML framework cover the mutants reported in published literature for in vitro/in vivo mutational studies on two case studies. The ML framework identifies the mutants for each of the three case studies reported in Section~\ref{Three_Case_studies}. Large difference in CA model parameter DCLVS between CoV-2 and CoV and (also CoV-2 and MERS) confirms difference in transmissibility of CoV-2 (2019) from that of CoV(2003) and MERS (2012). Further, CAML model also confirms same level of transmissibility for CoV-2 (2003) and MERS (2012). CAML platform can be employed to predict possible mutations in viral proteins. Six mutants predicted for CoV-2 are reported. We expect the results presented in our current and future papers will provide a platform to design therapeutic agents and robust vaccine with broader coverage to combat the CoV-2 epidemic. This epidemic is likely to persist for some more time due to new type of viral strains which may appear once the virus encounters vaccine administered on infected patients. Further, the CAML platform will provide an efficient workbench to develop therapeutic drug/vaccine for other types of viruses which may appear in future generating next pandemic situation.

\section*{References}
\bibliographystyle{elsarticle-num} 
\bibliography{References}

\begin{thebibliography}{10}
\expandafter\ifx\csname url\endcsname\relax
  \def\url#1{\texttt{#1}}\fi
\expandafter\ifx\csname urlprefix\endcsname\relax\def\urlprefix{URL }\fi
\expandafter\ifx\csname href\endcsname\relax
  \def\href#1#2{#2} \def\path#1{#1}\fi

\bibitem{bianchi2020sars}
M.~Bianchi, D.~Benvenuto, M.~Giovanetti, S.~Angeletti, M.~Ciccozzi,
  S.~Pascarella, Sars-cov-2 envelope and membrane proteins: structural
  differences linked to virus characteristics?, BioMed Research International
  2020.

\bibitem{sarkar2020structural}
M.~Sarkar, S.~Saha, Structural insight into the role of novel sars-cov-2 e
  protein: A potential target for vaccine development and other therapeutic
  strategies, PLoS One 15~(8) (2020) e0237300.

\bibitem{schoeman2020there}
D.~Schoeman, B.~C. Fielding, Is there a link between the pathogenic human
  coronavirus envelope protein and immunopathology? a review of the literature,
  Frontiers in microbiology 11 (2020) 2086.

\bibitem{de2020improved}
F.~De~Maio, E.~L. Cascio, G.~Babini, M.~Sali, S.~Della~Longa, B.~Tilocca,
  P.~Roncada, A.~Arcovito, M.~Sanguinetti, G.~Scambia, et~al., Improved binding
  of sars-cov-2 envelope protein to tight junction-associated pals1 could play
  a key role in covid-19 pathogenesis, Microbes and infection 22~(10) (2020)
  592--597.

\bibitem{kuo2007exceptional}
L.~Kuo, K.~R. Hurst, P.~S. Masters, Exceptional flexibility in the sequence
  requirements for coronavirus small envelope protein function, Journal of
  virology 81~(5) (2007) 2249--2262.

\bibitem{schoeman2019coronavirus}
D.~Schoeman, B.~C. Fielding, Coronavirus envelope protein: current knowledge,
  Virology journal 16~(1) (2019) 1--22.

\bibitem{kuzmin2021structure}
A.~Kuzmin, P.~Orekhov, R.~Astashkin, V.~Gordeliy, I.~Gushchin, Structure and
  dynamics of the sars-cov-2 envelope protein monomer, bioRxiv.

\bibitem{pervushin2009structure}
K.~Pervushin, E.~Tan, K.~Parthasarathy, X.~Lin, F.~L. Jiang, D.~Yu,
  A.~Vararattanavech, T.~W. Soong, D.~X. Liu, J.~Torres, Structure and
  inhibition of the sars coronavirus envelope protein ion channel, PLoS
  pathogens 5~(7) (2009) e1000511.

\bibitem{hassan2020sars}
S.~S. Hassan, P.~P. Choudhury, B.~Roy, Sars-cov2 envelope protein:
  non-synonymous mutations and its consequences, Genomics 112~(6) (2020)
  3890--3892.

\bibitem{wu2021effects}
S.~Wu, C.~Tian, P.~Liu, D.~Guo, W.~Zheng, X.~Huang, Y.~Zhang, L.~Liu, Effects
  of sars-cov-2 mutations on protein structures and intraviral protein--protein
  interactions, Journal of medical virology 93~(4) (2021) 2132--2140.

\bibitem{koyama2020variant}
T.~Koyama, D.~Platt, L.~Parida, Variant analysis of sars-cov-2 genomes,
  Bulletin of the World Health Organization 98~(7) (2020) 495.

\bibitem{westerbeck2019infectious}
J.~W. Westerbeck, C.~E. Machamer, The infectious bronchitis coronavirus
  envelope protein alters golgi ph to protect the spike protein and promote the
  release of infectious virus, Journal of virology 93~(11) (2019) e00015--19.

\bibitem{lorizate2011role}
M.~Lorizate, H.-G. Kr{\"a}usslich, Role of lipids in virus replication, Cold
  Spring Harbor perspectives in biology 3~(10) (2011) a004820.

\bibitem{maitra2020mutations}
A.~Maitra, M.~C. Sarkar, H.~Raheja, N.~K. Biswas, S.~Chakraborti, A.~K. Singh,
  S.~Ghosh, S.~Sarkar, S.~Patra, R.~K. Mondal, et~al., Mutations in sars-cov-2
  viral rna identified in eastern india: Possible implications for the ongoing
  outbreak in india and impact on viral structure and host susceptibility,
  Journal of Biosciences 45~(1) (2020) 1--18.

\bibitem{callaway2020making}
E.~Callaway, Making sense of coronavirus mutations, Nature 585 (2020) 174--177.

\bibitem{jenna1998effect}
S.~Jenna, C.~Sureau, Effect of mutations in the small envelope protein of
  hepatitis b virus on assembly and secretion of hepatitis delta virus,
  Virology 251~(1) (1998) 176--186.

\bibitem{cohen2011identification}
J.~R. Cohen, L.~D. Lin, C.~E. Machamer, Identification of a golgi
  complex-targeting signal in the cytoplasmic tail of the severe acute
  respiratory syndrome coronavirus envelope protein, Journal of virology
  85~(12) (2011) 5794--5803.

\bibitem{v2021coronavirus}
P.~V'kovski, A.~Kratzel, S.~Steiner, H.~Stalder, V.~Thiel, Coronavirus biology
  and replication: implications for sars-cov-2, Nature Reviews Microbiology
  19~(3) (2021) 155--170.

\bibitem{petersen2020comparing}
E.~Petersen, M.~Koopmans, U.~Go, D.~H. Hamer, N.~Petrosillo, F.~Castelli,
  M.~Storgaard, S.~Al~Khalili, L.~Simonsen, Comparing sars-cov-2 with sars-cov
  and influenza pandemics, The Lancet infectious diseases 20 (2020) e238--44.

\bibitem{cevik2020sars}
M.~Cevik, M.~Tate, O.~Lloyd, A.~E. Maraolo, J.~Schafers, A.~Ho, Sars-cov-2,
  sars-cov, and mers-cov viral load dynamics, duration of viral shedding, and
  infectiousness: a systematic review and meta-analysis, The lancet microbe 20.

\bibitem{kutter2018transmission}
J.~S. Kutter, M.~I. Spronken, P.~L. Fraaij, R.~A. Fouchier, S.~Herfst,
  Transmission routes of respiratory viruses among humans, Current opinion in
  virology 28 (2018) 142--151.

\bibitem{lee2010pdz}
H.-J. Lee, J.~J. Zheng, Pdz domains and their binding partners: structure,
  specificity, and modification, Cell communication and Signaling 8~(1) (2010)
  1--18.

\bibitem{zhang2020viral}
X.~Zhang, Y.~Tan, Y.~Ling, G.~Lu, F.~Liu, Z.~Yi, X.~Jia, M.~Wu, B.~Shi, S.~Xu,
  et~al., Viral and host factors related to the clinical outcome of covid-19,
  Nature 583~(7816) (2020) 437--440.

\bibitem{kawasuji2020transmissibility}
H.~Kawasuji, Y.~Takegoshi, M.~Kaneda, A.~Ueno, Y.~Miyajima, K.~Kawago,
  Y.~Fukui, Y.~Yoshida, M.~Kimura, H.~Yamada, et~al., Transmissibility of
  covid-19 depends on the viral load around onset in adult and symptomatic
  patients, PloS one 15~(12) (2020) e0243597.

\bibitem{venkatagopalan2015coronavirus}
P.~Venkatagopalan, S.~M. Daskalova, L.~A. Lopez, K.~A. Dolezal, B.~G. Hogue,
  Coronavirus envelope (e) protein remains at the site of assembly, Virology
  478 (2015) 75--85.

\bibitem{cai2018universal}
L.~Cai, H.~Bai, V.~Mahairaki, Y.~Gao, C.~He, Y.~Wen, Y.-C. Jin, Y.~Wang, R.~L.
  Pan, A.~Qasba, et~al., A universal approach to correct various hbb gene
  mutations in human stem cells for gene therapy of beta-thalassemia and sickle
  cell disease, Stem cells translational medicine 7~(1) (2018) 87--97.

\bibitem{hou2020sars}
Y.~J. Hou, S.~Chiba, P.~Halfmann, C.~Ehre, M.~Kuroda, K.~H. Dinnon, S.~R.
  Leist, A.~Sch{\"a}fer, N.~Nakajima, K.~Takahashi, et~al., Sars-cov-2 d614g
  variant exhibits efficient replication ex vivo and transmission in vivo,
  Science 370~(6523) (2020) 1464--1468.

\bibitem{seyran2020structural}
M.~Seyran, K.~Takayama, V.~N. Uversky, K.~Lundstrom, G.~Pal{\`u}, S.~P.
  Sherchan, D.~Attrish, N.~Rezaei, A.~A. Aljabali, S.~Ghosh, et~al., The
  structural basis of accelerated host cell entry by sars-cov-2, The FEBS
  journal.

\bibitem{hassan2020possible}
S.~Hassan, S.~Ghosh, D.~Attrish, P.~P. Choudhury, A.~A. Aljabali, B.~D. Uhal,
  K.~Lundstrom, N.~Rezaei, V.~N. Uversky, M.~Seyran, et~al., Possible
  transmission flow of sars-cov-2 based on ace2 features, Molecules 25~(24)
  (2020) 5906.

\bibitem{khailany2020genomic}
R.~A. Khailany, M.~Safdar, M.~Ozaslan, Genomic characterization of a novel
  sars-cov-2, Gene reports 19 (2020) 100682.

\bibitem{loo1994functional}
T.~W. Loo, D.~Clarke, Functional consequences of glycine mutations in the
  predicted cytoplasmic loops of p-glycoprotein., Journal of Biological
  Chemistry 269~(10) (1994) 7243--7248.

\bibitem{cabrera2021envelope}
D.~Cabrera-Garcia, R.~Bekdash, G.~W. Abbott, M.~Yazawa, N.~L. Harrison, The
  envelope protein of sars-cov-2 increases intra-golgi ph and forms a cation
  channel that is regulated by ph, The Journal of Physiology 599~(11) (2021)
  2851--2868.

\bibitem{kumar2021deletion}
B.~K. Kumar, A.~Rohit, K.~S. Prithvisagar, P.~Rai, I.~Karunasagar,
  I.~Karunasagar, Deletion in the c-terminal region of the envelope
  glycoprotein in some of the indian sars-cov-2 genome, Virus Research 291
  (2021) 198222.

\bibitem{zheng2021tlr2}
M.~Zheng, R.~Karki, E.~P. Williams, D.~Yang, E.~Fitzpatrick, P.~Vogel, C.~B.
  Jonsson, T.-D. Kanneganti, Tlr2 senses the sars-cov-2 envelope protein to
  produce inflammatory cytokines, Nature Immunology (2021) 1--10.

\bibitem{mukherjee2020host}
S.~Mukherjee, D.~Bhattacharyya, A.~Bhunia, Host-membrane interacting interface
  of the sars coronavirus envelope protein: Immense functional potential of
  c-terminal domain, Biophysical chemistry 266 (2020) 106452.

\bibitem{NatureEducation2008}
S.~Clancy, Genetic mutation, Nature Education 1~(1) (2008) 187.

\bibitem{ncbiDatabase}
\href{https://www.ncbi.nlm.nih.gov/refseq/}{Refseq: Ncbi reference sequence
  database}.
\newline\urlprefix\url{https://www.ncbi.nlm.nih.gov/refseq/}

\bibitem{von1996theory}
J.~Von~Neumann, A.~W. Burks, Theory of self-reproducing automata, University of
  Illinois Press Urbana, 1996.

\bibitem{wolfram1983statistical}
S.~Wolfram, Statistical mechanics of cellular automata, Reviews of modern
  physics 55~(3) (1983) 601--644.

\bibitem{codd1968cellular}
E.~F. Codd, Cellular Automata, Academic Press Inc, New York, 1968.

\bibitem{conway1970game}
J.~Conway, The game of life, Scientific American.

\bibitem{wolfram2002new}
S.~Wolfram, A new kind of science, Wolfram-Media Inc., Champaign, Ilinois,
  United States, 2002.

\bibitem{chaudhuri2018new}
P.~P. Chaudhuri, S.~Ghosh, A.~Dutta, S.~P. Choudhury, A New Kind of
  Computational Biology: Cellular Automata Based Models for Genomics and
  Proteomics, Springer Nature Singapore, ISBN 978-981-1-13-163, 2018.

\bibitem{ppc1}
P.~Pal~Chaudhuri, D.~Roy~Chowdhury, S.~Nandi, S.~Chattopadhyay, Additive
  Cellular Automata -- Theory and Applications, Vol.~1, IEEE Computer Society
  Press, Los Alamitos, USA, ISBN 0-8186-7717-1, 1997.

\bibitem{jithesh2021model}
P.~Jithesh, A model based on cellular automata for investigating the impact of
  lockdown, migration and vaccination on covid-19 dynamics, Computer methods
  and programs in biomedicine 211 (2021) 106402.

\bibitem{cavalcante2021modelling}
A.~L.~B. Cavalcante, L.~P. de~Faria~Borges, M.~A. da~Costa~Lemos, M.~M.
  de~Farias, H.~S. Carvalho, Modelling the spread of covid-19 in the capital of
  brazil using numerical solution and cellular automata, Computational biology
  and chemistry 94 (2021) 107554.

\bibitem{schimit2021model}
P.~H. Schimit, A model based on cellular automata to estimate the social
  isolation impact on covid-19 spreading in brazil, Computer Methods and
  Programs in Biomedicine 200 (2021) 105832.

\bibitem{ghosh2020data}
S.~Ghosh, S.~Bhattacharya, A data-driven understanding of covid-19 dynamics
  using sequential genetic algorithm based probabilistic cellular automata,
  Applied Soft Computing 96 (2020) 106692.

\bibitem{pereira2021deep}
F.~H. Pereira, P.~H. Schimit, F.~E. Bezerra, A deep learning based surrogate
  model for the parameter identification problem in probabilistic cellular
  automaton epidemic models, Computer Methods and Programs in Biomedicine 205
  (2021) 106078.

\bibitem{ridley2015evolution}
M.~Ridley, The Evolution of Everything: How Small Changes Transform Our World,
  Harper Collins UK, ISBN 978-0-00-754247-5, 2015.

\bibitem{hancock2008detecting}
A.~M. Hancock, A.~Di~Rienzo, Detecting the genetic signature of natural
  selection in human populations: models, methods, and data, Annual review of
  anthropology 37 (2008) 197--217.

\bibitem{vitti2013detecting}
J.~J. Vitti, S.~R. Grossman, P.~C. Sabeti, Detecting natural selection in
  genomic data, Annual review of genetics 47 (2013) 97--120.

\bibitem{voight2006map}
B.~F. Voight, S.~Kudaravalli, X.~Wen, J.~K. Pritchard, A map of recent positive
  selection in the human genome, PLoS biology 4~(3) (2006) e72.

\bibitem{carlice2017investigation}
T.~Carlice-dos Reis, J.~Viana, F.~C. Moreira, G.~d.~L. Cardoso, J.~Guerreiro,
  S.~Santos, A.~Ribeiro-dos Santos, Investigation of mutations in the hbb gene
  using the 1,000 genomes database, PloS one 12~(4) (2017) e0174637.

\bibitem{hu2020comparison}
T.~Hu, Y.~Liu, M.~Zhao, Q.~Zhuang, L.~Xu, Q.~He, A comparison of covid-19, sars
  and mers, PeerJ 8 (2020) e9725.

\bibitem{zhu2020sars}
Z.~Zhu, X.~Lian, X.~Su, W.~Wu, G.~A. Marraro, Y.~Zeng, From sars and mers to
  covid-19: a brief summary and comparison of severe acute respiratory
  infections caused by three highly pathogenic human coronaviruses, Respiratory
  research 21~(224) (2020) 1--14.

\end{thebibliography}

\end{document}